\documentclass[sn-mathphys,Numbered,onecolumn]{sn-jnl}% Math and Physical Sciences Reference Style
\usepackage{graphicx}%
\usepackage{multirow}%
\usepackage{amsmath,amssymb,amsfonts}%
\usepackage{amsthm}%
\usepackage{mathrsfs}%
\usepackage[title]{appendix}%
\usepackage{xcolor}%
\usepackage{textcomp}%
\usepackage{manyfoot}%
\usepackage{booktabs}%
\usepackage{algorithm}%
\usepackage{algorithmicx}%
\usepackage{algpseudocode}%
\usepackage{listings}%
\theoremstyle{thmstyleone}%
\theoremstyle{thmstyletwo}%

\theoremstyle{thmstylethree}%

\newcommand{\ExoSim}{\texttt{ExoSim~2}}

\usepackage{aas-macros}

\begin{document}

\title[ExoSim~2]{ExoSim~2: the new Exoplanet Observation Simulator applied to the Ariel space mission}


\author*[1,2,3,4]{\fnm{Lorenzo V.} \sur{Mugnai}}\email{ mugnail@cardiff.ac.uk}
\author[1]{\fnm{Andrea} \sur{Bocchieri}}
\author[1]{\fnm{Enzo} \sur{Pascale\\}}
\author[5]{\fnm{Andrea} \sur{Lorenzani}}
\author[2]{\fnm{Andreas} \sur{Papageorgiou}}

\affil*[1]{\orgdiv{Dipartimento di Fisica}, \orgname{La Sapienza Universit\`a di Roma}, \orgaddress{\street{Piazzale Aldo Moro 5}, \city{Roma}, \postcode{00185}, \country{Italy}}}

\affil[2]{\orgdiv{School of Physics and Astronomy}, \orgname{Cardiff University}, \orgaddress{\street{Queens Buildings, The Parade}, \city{Cardiff}, \postcode{CF24 3AA}, \country{UK}}}

\affil[3]{\orgdiv{Department of Physics and Astronomy}, \orgname{University College London}, \orgaddress{\street{Gower Street}, \city{London}, \postcode{WC1E 6BT}, \country{UK}}}

\affil[4]{\orgdiv{Osservatorio Astronomico di Palermo}, \orgname{INAF}, \orgaddress{\street{Piazza del Parlamento 1}, \city{Palermo}, \postcode{90134}, \country{Italy}}}
\affil[5]{\orgdiv{Osservatorio Astrofisico di Arcetri}, \orgname{INAF}, \orgaddress{\street{Largo E. Fermi 5}, \city{Firenze}, \postcode{50125}, \country{Italy}}}


\abstract{ExoSim~2 is the next generation of the Exoplanet Observation Simulator (ExoSim) tailored for spectro-photometric observations of transiting exoplanets from space, ground, and sub-orbital platforms. This software is a complete rewrite implemented in Python 3, embracing object-oriented design principles, which allow users to replace each component with their functions when required. ExoSim~2 is publicly available on GitHub, serving as a valuable resource for the scientific community.
ExoSim~2 employs a modular architecture using Task classes, encapsulating simulation algorithms and functions. This flexible design facilitates the extensibility and adaptability of ExoSim~2 to diverse instrument configurations to address the evolving needs of the scientific community. Data management within ExoSim~2 is handled by the Signal class, which represents a structured data cube incorporating time, space, and spectral dimensions.
The code execution in ExoSim~2 follows a three-step workflow: the creation of focal planes, the production of Sub-Exposure blocks, and the generation of non-destructive reads (NDRs). Each step can be executed independently, optimizing time and computational resources.
ExoSim~2 has been extensively validated against other tools like ArielRad and has demonstrated consistency in estimating photon conversion efficiency, saturation time, and signal generation. The simulator has also been validated independently for instantaneous read-out and jitter simulation, and for astronomical signal representation.
In conclusion, ExoSim~2 offers a robust and flexible tool for exoplanet observation simulation, capable of adapting to diverse instrument configurations and evolving scientific needs. Its design principles and validation results underscore its potential as a valuable resource in the field of exoplanet research.}




\maketitle

\section{Introduction}
\label{Intro} 

The blossoming field of exoplanets, planets beyond our solar system, stands as a revolutionary chapter in the annals of astrophysics. Transit photometry and slit-less low-resolution spectroscopy from the visible to the mid-IR region of the electromagnetic spectrum are significantly bolstered by observational simulators like \texttt{ExoSim} \citep{Sarkar2021}. These simulators produce synthetic data reflecting the potential outputs of observing facilities, thus enabling the evaluation and validation of observational methodologies, noise models, known systematics, and data analysis techniques, even before the actual data collection commences. Serving as a comprehensive testbed, this synthetic data empowers researchers to refine and optimize the performance of observational tools, prototype data pipelines, and retrieval algorithms. Hence, it significantly contributes to the judicious use of telescope time and resources \citep{Batalha2017PASP}.

Exoplanetary science has found particularly impactful applications of observational simulators in optimizing the characterization of planetary atmospheres. Leveraging synthetic data, researchers can investigate a broad spectrum of atmospheric conditions. This methodical exploration of diverse atmospheric features allows for investigating the impact of varying chemical compositions, temperature-pressure profiles, and cloud properties on the detectability of atmospheric signatures \citep{Mugnai2021, Changeat2020, Bocchieri2023}. The multifaceted insights gained thus facilitate the refinement of retrieval techniques while also providing foresight into potential challenges tied to atmospheric characterization, hence, guiding future observational campaigns \citep{Greene2016}.

The pivotal role of observational simulators in exoplanetary science extends to the realm of mission proposal planning and implementation. Simulated outputs offer proof-of-concept demonstrations of proposed observational strategies and reveal the scientific potential of mission ideas \citep{Kaltenegger2009}. By providing a quantitative backbone for mission design and expected outcomes, simulators have become indispensable in securing funding and telescope time for the expanding domain of exoplanet discovery and characterization.

The Planetary Transits and Oscillations of stars (PLATO) Simulator is a manifestation of such a tool, with proven efficacy in transit detection contexts. This simulator generates synthetic light curves, encompassing various instrumental effects and stellar properties, to evaluate transit detection and false positive rejection algorithms \citep{Marcos-Arenal2014}. Although it provides a detailed portrayal of transit detections, its utility is confined to the specificity of the PLATO mission \citep{Rauer2014}, and extensive modification is necessary for integration with different observational campaigns. Moreover, the underlying assumptions in its stellar and noise models may introduce inconsistencies when juxtaposed with real observations.

In the context of the James Webb Space Telescope (JWST) \citep{Greene2016}, the PandExo simulator stands out as a significant tool \citep{Batalha2017PASP}. Crafted specifically for JWST exoplanet observations, PandExo provides precise pixel-level signal-to-noise ratio (SNR) estimates for proposed observations, thereby aiding JWST observation planning and data analysis strategy development. Similar to the PLATO simulator, the utility of PandExo is intrinsically tied to JWST, and its reliability is subject to assumptions about instrument performance.
Meanwhile, ExoNoodle presents itself as a versatile, mission-independent tool \citep{Martin-Lagarde2021}. It simulates photometric and radial velocity observations associated with star-planet systems, thus enabling integration with instrument simulators such as MIRIsim \citep{Geers2019}.

% Terminus represents another evolution in exoplanet observation simulators, providing a means to model not only the observational data from space-based telescopes but also the atmospheric properties of the observed exoplanets. Despite its extensive feature set and versatility, the application and interpretation of Terminus outputs necessitate an advanced understanding of exoplanetary atmospheres and observational techniques \citep{Edwards2021}.

The Atmospheric Remote-sensing Infrared Exoplanet Large-survey (Ariel) is a mission spearheaded by the European Space Agency (ESA) that aims to delve into the nature of exoplanet atmospheres \citep{Tinetti2018}. With a projected launch towards the end of the 2020 decade, Ariel will observe transits and eclipses of approximately 1,000 exoplanets spanning various sizes, temperatures, and host star characteristics \citep{Edwards2019}. By producing a statistically significant sample of exoplanet atmospheres, Ariel aspires to answer fundamental questions about the genesis and evolution of these celestial bodies.

In anticipation of the Ariel mission, simulators like \texttt{ArielRad} \citep{Mugnai2020} and \texttt{ExoSim} \citep{Sarkar2021} have been developed to model expected spacecraft observations, using the heritage left by \texttt{EChoSim} \citep{EChOSim} for the EChO mission \citep{EChO}. \texttt{ArielRad} is a radiometric model that simulates the total system efficiency of the Ariel payload, from the entrance pupil to the detector. This tool provides crucial insight into the expected performance of Ariel instruments, includes potential systematic uncertainties, and predicts data quality that Ariel is expected to obtain, thereby enabling studies to optimize the mission's observing strategy. \texttt{ArielRad} is based on the generic radiometric simulator, \texttt{ExoRad}\footnote{\url{https://github.com/ExObsSim/ExoRad2-public}} \citep{ExoRad, ExoRad.ascl}, which is publicly available.

\texttt{ExoSim} \citep{Sarkar2021}, a simulator for exoplanet observations, generates synthetic observational data, representative of the expected outputs of space-based detection missions, including Ariel. It simulates diverse noise sources such as photon noise, instrumental noise, and zodiacal light, providing robust estimates of achievable precision in exoplanet observations. \texttt{ExoSim} serves multiple purposes, including:
\begin{enumerate}
	\item informing instrumental and mission design;
	\item evaluating instrumental and astrophysical systematics;
	\item validating the mission's compatibility with scientific objectives and evaluating the associated margins;
	\item testing and fine-tuning data analysis techniques;
	\item assisting in the selection of exoplanetary targets for observation.
\end{enumerate}

Both \texttt{ArielRad/ExoRad} and \texttt{ExoSim} play pivotal roles in mission preparation and algorithmic development, thereby fortifying the prospects for successful mission execution and subsequent data analysis and interpretation. While \texttt{ExoRad} is specifically tailored to simulate the radiometric performance, \texttt{ExoSim} provides comprehensive simulations of the observational data, incorporating time-dependent systematics. \texttt{ExoSim} and \texttt{ExoRad} have gained significant recognition and adoption within the scientific community, with \texttt{ExoSim} being adapted for other missions such as Twinkle \citep{Twinkle}, EXCITE \citep{Excite}, and JWST \citep{jexosim, JexoSim2} through dedicated versions of the code.

Despite the invaluable insights provided by observation simulators, like ExoSim, their complexity presents a unique set of challenges. Development and execution require considerable time and an in-depth understanding of both the simulation process and the intricate instrument designs. The resulting complexity restricts simulator applications, in many cases confining their use to highly refined designs and limiting their deployment to a narrow range of scenarios.

Addressing these limitations necessitates three key improvements. Firstly, the development of flexible simulators that can adapt to new observatories, providing seamless transitions to new observational platforms. Secondly, the creation of intuitive and user-friendly interfaces to increase accessibility and stimulate broader usage, thereby fuelling the potential for new discoveries. Lastly, it will ensure growth capacity within these simulators to accommodate new features and advancements, maintaining their relevance in the rapidly evolving field of exoplanet observations.

Meeting these demands not only streamlines the adoption of these simulators by a wider user base but also extends their potential use cases, thereby accelerating the pace of discovery in the exoplanet field. Thus, a solution would be to iterate on the existing tools to find the best compromise between complexity and wide applicability. It is in this spirit that we developed \ExoSim, aiming to present the scientific community with a definitive tool that can address their evolving needs.

\section{The Simulator}
\label{simulator} 

\ExoSim\ is an upgraded version of \texttt{ExoSim} \citep{Sarkar2021}, implemented in Python 3 with a focus on object-oriented design principles. The source code is publicly available on GitHub\footnote{\url{https://github.com/arielmission-space/ExoSim2-public}} and can be installed from PyPI\footnote{\url{https://pypi.org/project/exosim/}} under the BSD3-Clause License. For this study, we utilize version \texttt{2.0.0-rc2}\footnote{\url{https://github.com/arielmission-space/ExoSim2-public/releases/tag/v2.0.0-rc2}}, which has undergone extensive testing on Linux and iOS systems, supporting Python 3.8, 3.9, and 3.10, and achieving a test coverage exceeding 90\%.

The primary input for \ExoSim\ is an \texttt{.xml} file, allowing users to define simulation parameters and reference necessary data files. Comprehensive instructions for filling the \texttt{.xml} file are available in the accompanying software documentation\footnote{\label{documentation}\url{http://exosim2-public.readthedocs.io/}}.

\ExoSim\ is designed with a modular architecture that utilizes \texttt{Task} classes to encapsulate various simulation algorithms and functions. While the code comes with a set of default \texttt{Task}s, it also offers the flexibility for users to create custom \texttt{Task}s that meet specific simulation needs. In essence, nearly every component of \ExoSim\ can be replaced with a user-defined function, thanks to its \texttt{Task}-based structure. The documentation offers in-depth explanations of the default \texttt{Task}s and provides comprehensive guidelines for developing and implementing custom \texttt{Task}s. This design flexibility ensures that \ExoSim\ can be easily adapted to accommodate a wide range of instrument configurations. We emphasize that the \texttt{Task} structure allows \ExoSim\ to be highly extensible, and capable of adapting to the specific requirements of different telescopes.

\begin{figure}
	\centering
	\includegraphics[width=0.8\linewidth]{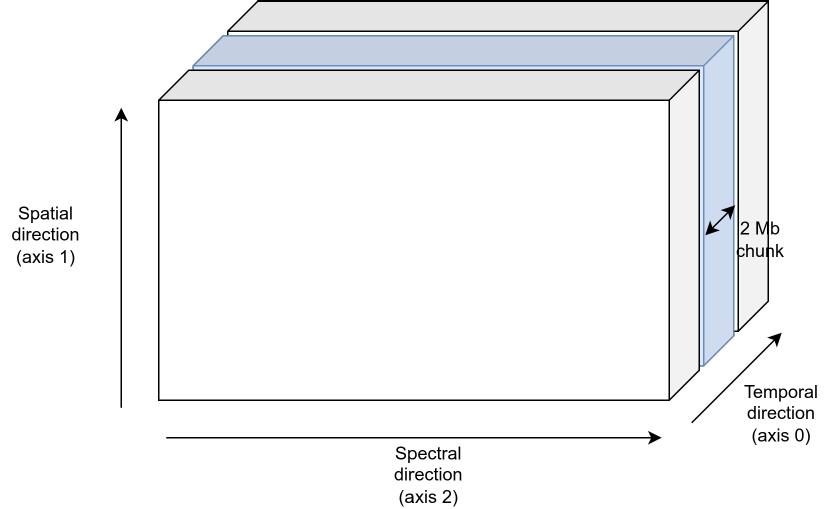}
	\caption{Data cubes based on the \texttt{Signal} class are employed within \ExoSim\ for data manipulation. These cubes have three axes: time (axis 0), space (axis 1), and spectral direction (axis 2). When the data cube encompasses Sub-Exposure or NDRs generated by \ExoSim, the \texttt{Signal} class can utilize the HDF5 chunking system to partition it into temporal slices, enabling efficient storage and handling of large data volumes ($>$10 Gbs).}
	\label{fig:signal_class}
\end{figure}

Data management within \ExoSim\ is handled by the \texttt{Signal} class, which represents a structured data cube incorporating time, space, and spectral dimensions (see Fig. \ref{fig:signal_class}). The \texttt{Signal} class efficiently manages metadata associated with the data, including temporal, spatial, and spectral axes, while offering essential functionalities for performing common data operations. For more detailed information, please consult the provided code documentation.

\begin{figure}
	\centering
	\includegraphics[width=0.8\linewidth]{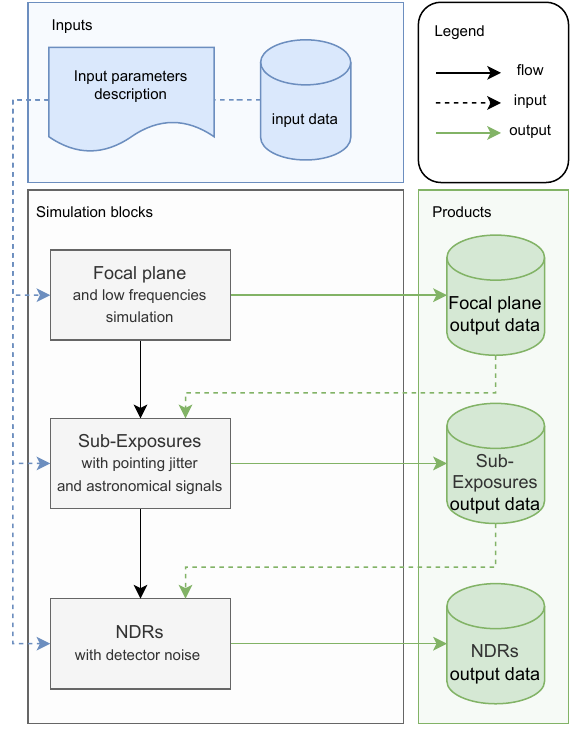}
	\caption{\ExoSim\ operates via a three-step workflow, where each step can be executed independently. The initial step involves the creation of focal planes based on source and instrument specifications. Subsequently, the Sub-Exposure blocks simulate high-frequency sampled focal planes, incorporating astronomical signals, pointing jitter, and detector read-out modes. Finally, the NDR block generates non-destructive reads (NDRs) from the Sub-Exposure, accounting for detector noise.}
	\label{fig:exosimblocks}
\end{figure}

The code execution in \ExoSim\ follows a three-step workflow (see Fig. \ref{fig:exosimblocks}). Each step can be executed independently, allowing users to explore different configurations without re-running the entire simulation, thereby optimizing time and computational resources. The first step involves the creation of focal planes (see Section \ref{focal plane creation} for details). This process generates time-dependent focal planes for each instrument channel, sampled at low frequencies corresponding to hour timescales. Subsequently, the Sub-Exposure blocks (Section \ref{sub-exposure}) produce high-frequency sampled focal planes (fraction-of-a-second timescales) while accounting for various factors such as pointing jitter and detector read-out mode. Finally, the NDR block (Section \ref{NDRs}) generates non-destructive reads (NDRs) from the Sub-Exposure, incorporating detector noise and jitter into the simulated data.

In the subsequent sections, we illustrate the simulation workflow by simulating an observation of HD 209458 with the Ariel space mission. We selected this target as it is a standard reference for the mission \citep{Mugnai2020}. We adopt the mission parameters from the B2-phase\footnote{The parameters used exhibit slight variations compared to those presented in \cite{Mugnai2020} from the B1-phase}. The payload configurations used in this study are stored in a dedicated GitHub repository managed by the Ariel mission, with restricted access to members with key responsibilities to foster reproducibility without divulging restricted information.

\subsection{Focal Plane creation}
\label{focal plane creation}
The focal plane creation block returns a data cube containing the focal plane images vs. time. The images are sampled at the pixel or sub-pixel level, depending on whether the oversampling factor, $osf$, is 1 or larger. Sampling the detector at the sub-pixel level assures an adequate sampling of the jitter movements for the optimized jitter simulation implemented by \ExoSim\ and described in Sec. \ref{sub-exposure}. 
This parameter is set by the user in the configuration file: if the detector array is made up of $N \times M$ physical pixels and is sampled in $t_{lf}$ time steps, where $lf$ is for ``low frequencies'', the resulting focal plane images will be a matrix of $(N \cdot osf) \times (M \cdot osf) \times t_{lf}$ elements. 
For brevity, we omit in the following that all the quantities (signals, efficiencies, etc.) considered in Sec. \ref{focal plane creation} and \ref{sub-exposure} can be time-dependent (sampled at $t_{lf}$).    

\begin{figure}
	\centering
	\includegraphics[width=0.6\linewidth]{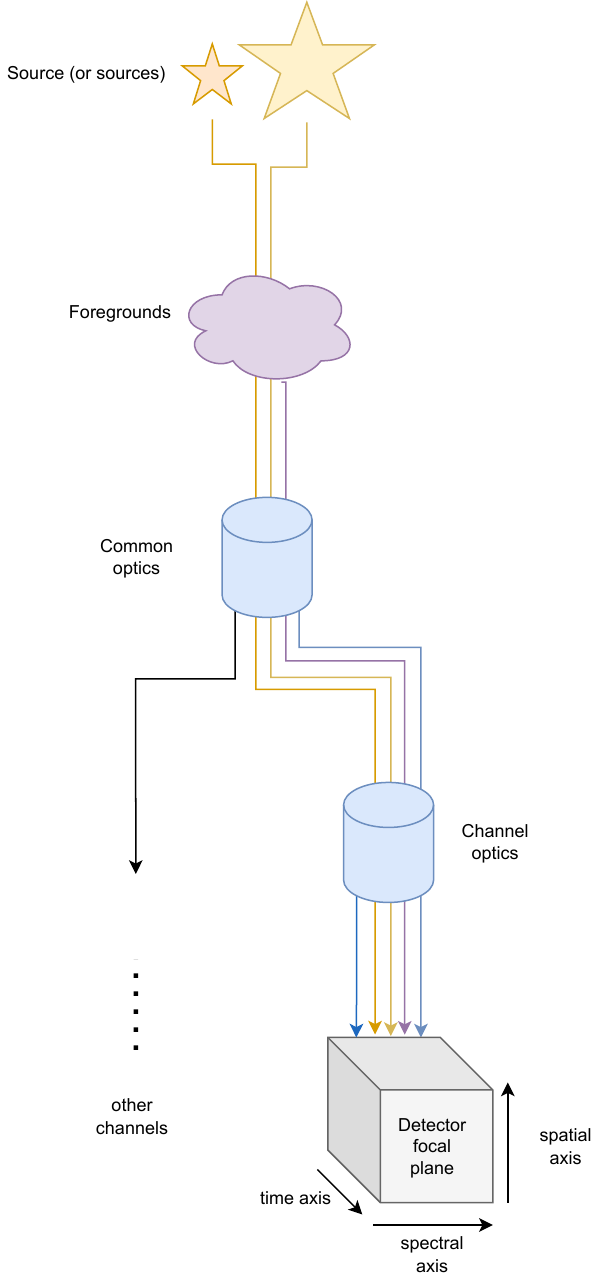}
	\caption{To reach the focal plane, the light from each source is first filtered by the various foregrounds and then passes through the optical chain of the system. Each element in the chain transmits the incoming light and emits its own radiation (here represented by the colour-coded arrows). The simulator can handle multiple focal planes on the same instrument, called ``channels''. After the common optical path, the light is split between channels to reach the respective focal plane. The final representation of the light collected by each focal plane is a data cube with spatial, spectral, and temporal directions (see Fig. \ref{fig:signal_class}).}
	\label{fig:roadtofocalplane}
\end{figure}

Fig. \ref{fig:roadtofocalplane} reports a diagram with the path followed by the light propagating from a generic point source to the detector focal plane. The current \ExoSim\ version supports only point sources. Diffuse sources and imaging support will be added in the successive code versions. Also, \ExoSim\ supports multiple sources in the field, located in the focal plane according to their position in the sky and the telescope pointing direction. As described in Sec. \ref{final-focal-plane} if multiple sources are provided \ExoSim\ produces two separated focal planes, following the same procedure, separating the science target of the observation from the other stars.

\subsubsection{Target Source \label{sec:target_source}}

Following \cite{Mugnai2020, Sarkar2021, jexosim, JexoSim2}, the Spectral Energy Density (SED) of the target star  at the telescope input is evaluated as 
\begin{equation}
	S(\lambda) = \frac{R_\star^2}{D^2} S_\star(\lambda)
\end{equation}
where $R_\star$ represents the radius of the host star, $D$ is the distance and $S(\lambda_\star)$ denotes the SED of the star. The stellar SED is interpolated from a grid of synthetic Phoenix spectra\footnote{\url{https://phoenix.ens-lyon.fr/Grids/BT-Settl/CIFIST2011_2015/FITS/}} \citep{Baraffe2015} and the best matching model is selected based on stellar temperature, surface gravity, and metallicity. The resulting SED has units of $W / (m^2    \cdot \mu m)$.
To represent the expected spectrum of the star HD 209458 we used the Phoenix file \texttt{lte061.0-4.5-0.0a+0.0.BT-Settl.spec.fits}.

\begin{figure}
	\centering
	\includegraphics[width=\linewidth]{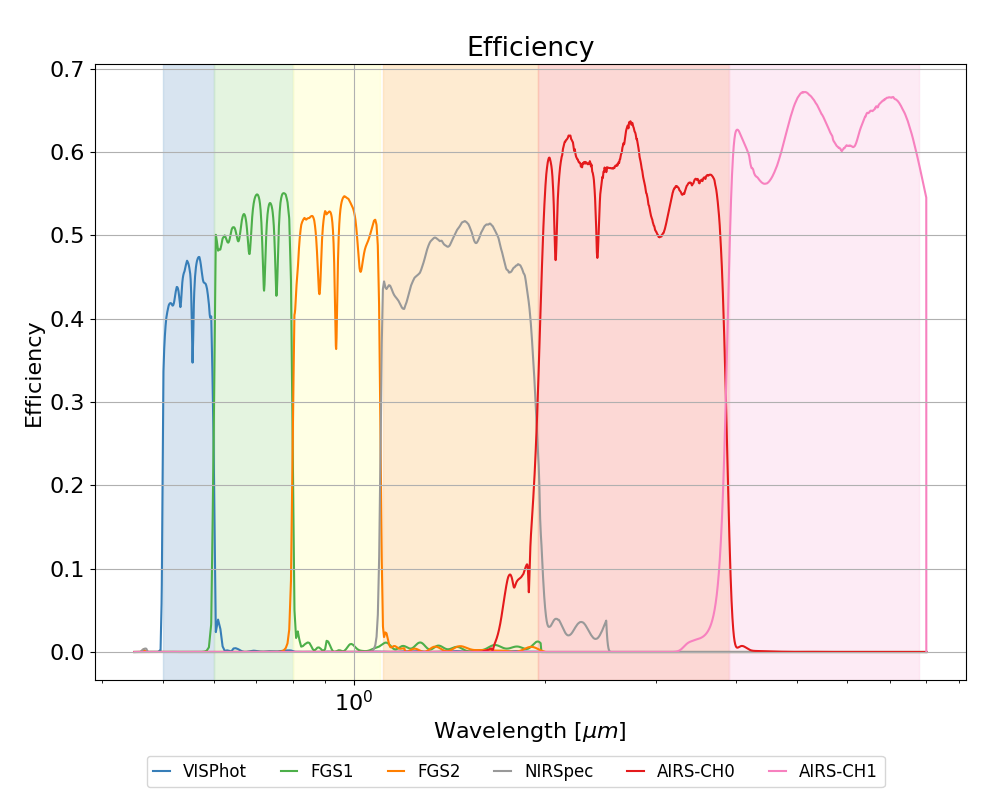}
	\caption{Total transmission for the optical paths of Ariel's channels.}
	\label{fig:efficiency}
\end{figure}

The light is then scaled by the transmission of each optical element so that the total wavelength-dependent transmission of the optical path $\Phi (\lambda)$ is
\begin{equation}
	\Phi (\lambda) = \prod_i \phi_{i}(\lambda)
\end{equation}
where $\phi_{i}(\lambda)$ refers to the wavelength-dependent transmission of the $i$-th optical element, including eventual foreground that can also be of astrophysical origin such as Zodiacal light, exo-Zodiacal light, Galactic extinction, etc. The total instrument transmission for the channels of the Ariel space mission is depicted in Fig. \ref{fig:efficiency}.

Next, the stellar SED is multiplied by the average wavelength-dependent detector responsivity, $\nu(\lambda)$, which has units of $counts/J$, and represents the detector ability to convert the incoming energy into elementary charges in the pixel, and by the telescope aperture $A_{tel}$, to obtain the flux density, $FD$, in units of $counts/(s \cdot \mu m)$. Finally, the stellar flux, in units of $counts/s$, is obtained by binning the flux density on the detector pixel or sub-pixel wavelength grid. In the case of a spectrometer, \ExoSim\ computes such a grid from the wavelength solution. The $i$-th pixel collects the light between $\lambda_{x_i, \, min}$ and $\lambda_{x_i, \, max}$ on the $x$ direction and $\lambda_{y_i, \, min}$ and $\lambda_{y_i, \, max}$ on the $y$ direction. Therefore, this recipe supports curved wavelength solutions for spectrometers.
In a photometer, every pixel is sensitive to the same wavelength range. 
To compute the flux measured by the $i$-th pixel, \ExoSim\ convolves every sampled wavelength with a monochromatic Point-Spread-Function (PSF) estimating the contribution of every wavelength to each pixel:

\begin{equation}
	\begin{split}
		F(x_i, y_i) =& A_{tel} \int_{\lambda} \Phi(\lambda) \cdot \nu(\lambda) \cdot S(\lambda)\\
		& \cdot PSF(u(\lambda)-x_i,v(\lambda)-y_i, \lambda) \, \textrm{d} \lambda   
	\end{split}
\end{equation}
where the relation between $u$, $v$, and $\lambda$ is described by the linear dispersion law between wavelength and position along the $x$ and $y$ axis respectively.

\begin{figure*}	
	\centering
	\includegraphics[width=1\linewidth]{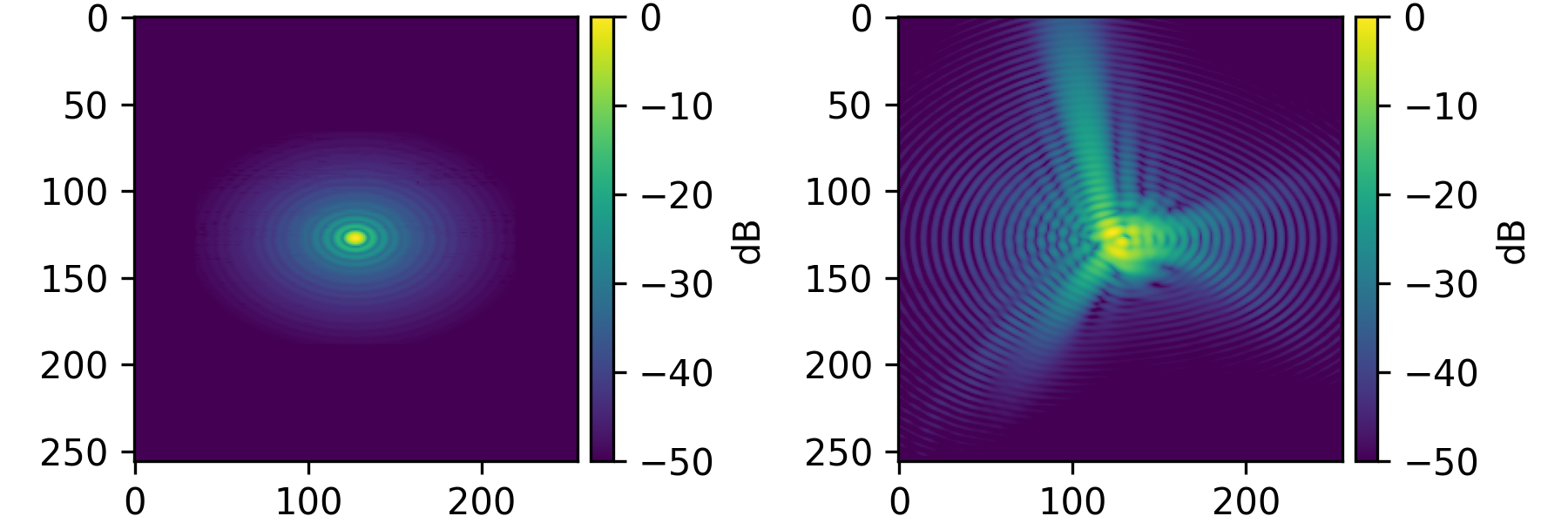}
	\caption{Focal planes produced by \ExoSim\ for the Ariel VISPhot channel using two different PSFs. The focal plane array size is $64 \times 64$ pixels, oversampled by a factor of 3. Both panels are presented in decibels ($[dB]$) units to highlight the PSF side lobes. The left panel shows the effective PSF using an Airy function, which is cropped to the 8th zero. The right panel presents the effective PSF using a PAOS-simulated PSF as input, including aberrations and a defocusing of $200 \, \text{nm}$ root mean square (RMS) on the wavefront.}% It should be noted that for the Ariel mission goals, the photometers are used as ``light buckets'', so the specific shape and aberrations of the PSF are not critical as long as the PSF is Nyquist-sampled and the flux is collected in a sufficiently compact region of the focal plane.}
\label{fig:focalplanePSFs}
\end{figure*}

\ExoSim\ supports diffraction PSFs (Airy) or Gaussian functions. Alternatively, custom PSFs can be loaded from a file. \texttt{PAOS} \citep{PAOS} products are natively supported. A depiction of an aberrated \texttt{PAOS} PSF and an Airy PSF is shown in Figure \ref{fig:focalplanePSFs}.

The PSFs are typically normalized to unity in volume; however, they can be normalized to a value smaller than unity if there is unaccounted power loss in the optical transmission $\Phi (\lambda)$. For a spectrometer, \ExoSim\ selects the correct monochromatic PSF, sampled by the $i$-th pixel, multiplies it by the corresponding source flux $F_i$, and adds the result to the focal plane. 
In a photometer, \ExoSim\ samples the channel wavelength range with monochromatic PSFs, multiplies them by the corresponding source flux, and sums them on the focal plane to obtain the effective PSF.

\subsubsection{Diffuse light: foregrounds and optical paths}
A foreground in \ExoSim\ is represented by an optical element positioned between the source and the telescope, generating diffuse light as opposed to a point source. \ExoSim\ supports custom foregrounds, allowing users to define their own specifications. The properties of each foreground are described by wavelength-dependent transmission and radiance functions. If the input file includes multiple foregrounds, \ExoSim\ processes them sequentially, considering the ordering in the configuration file, as each foreground contributes solely to the transmission of the fraction of the incident light that reaches it (refer to Fig. \ref{fig:roadtofocalplane}).

% \begin{figure}
% 	\centering
% 	\includegraphics[width=1\linewidth]{foregrounds.png}
% 	\caption{Multiple foregrounds can be included in the simulation. According to the order they are listed in the configuration file, each foreground transmits part of the incoming light and produces its emission.}
% 	\label{fig:foregrounds}
% \end{figure}

\ExoSim\ incorporates the Zodiacal foreground as a default foreground. The zodiacal radiance is modeled in a similar fashion to the methodology described in \cite{Mugnai2020}. We utilize a modified version of the JWST-MIRI Zodiacal model \citep{Glasse2010} scaled based on the target's position in the sky, fitted to the model presented in \citep{Kelsall1998}:

\begin{equation}
\begin{split}
	I_{zodi}(\lambda) =& A\, ( 3.5 \times 10^{-14} \, BB(\lambda, 5500 \, K) \\ 
	&+ 3.58 \times 10^{-8} \, BB(\lambda, 270 \, K) )
\end{split}
\end{equation}
where the coefficient $A$ is defined in the configuration file or determined by fitting it to the pointed direction in the sky, and $\text{BB}$ represents the Black Body radiance calculated for specific temperatures. The foreground radiance is subsequently scaled by the field of view of the pixels, $\Omega$, which is defined in \ExoSim\ following the analytical solution for the solid angle subtended by an ellipse reported in Eq. 56 by \cite{Conway2010}.

% \begin{equation}
% \begin{split}
	% \Omega(F\#_M, F\#_m) =& 2 \pi - 8\frac{F\#_m}{F\#_M}\sqrt{\frac{4 F\#_m^2}{F\#_m^2 +1}} \\
	% &\Pi \left( \sqrt{1-\frac{F\#^2_m}{F\#_M^2}}, \, \sqrt{\frac{F#_M^2-F#_m^2}{F\#_M^2(F\#_m^2 +1)}} \right)
	% \end{split}
% \end{equation}
% where $\Pi$ is the complete elliptic integral of the third kind.

Upon reaching a telescope, the light interacts with its optical elements. Similarly to the foregrounds, each optical element transmits a portion of the incoming light based on its wavelength-dependent efficiency and emits radiation based on its wavelength-dependent emissivity. The radiance of the optical element is typically modeled as a Black Body function scaled by the wavelength-dependent emissivity. 
Most modern telescopes have a common optical path prior to splitting the light into different instruments, referred to as ``channels'' in \ExoSim. The \ExoSim\ code allows for the simultaneous simulation of multiple channels with their respective optical paths.

\begin{figure}
\centering
\includegraphics[width=0.8\linewidth]{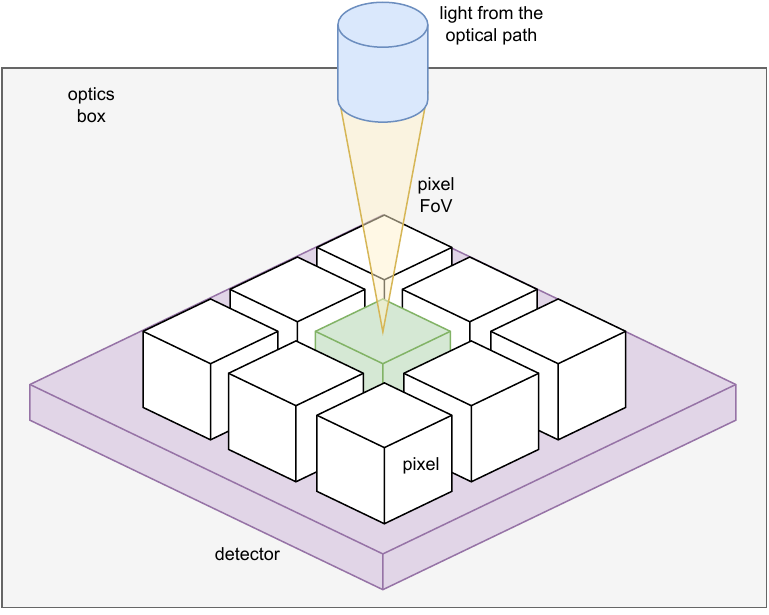}
\caption{Each pixel on the focal plane receives distinct light contributions. The illumination of a pixel due to the light from the optical path can be described by the pixel solid angle $\Omega$. The radiation originating from the optical elements located in front of the pixel contributes from an angle of $\pi - \Omega$, whereas the detector box positioned behind the pixel irradiates from an angle of $\pi$. All these contributions are accounted for in the simulations.}
\label{fig:detectorirradiation}
\end{figure}

As with the foregrounds, the radiance $I(\lambda)$ of the optical element is multiplied by the pixel field of view when it reaches the focal plane. Among the simulated optical elements are the optics box and the detector box. These represent the environment containing the optical elements and the detector, respectively. When light from the optical path irradiates each pixel, it does so according to the pixel's field of view. However, photons from the optics box and the detector box illuminate the pixel from different solid angles. As illustrated in Fig. \ref{fig:detectorirradiation}, the light from the optics box irradiates the pixel from above, excluding the pixel's field of view, and its radiance is scaled by $\pi - \Omega$, where again $\Omega$ is the pixel solid angle. On the other hand, the detector box irradiates the pixel from behind, and its radiance is scaled by $\pi$. 

It's worth noting that users have the flexibility to define the solid angle $\omega$ in steradian units for different optical elements via the configuration file. To further tailor the simulation, users can also create custom functions to describe each optical element, allowing for a more accurate representation of wavelength and time-dependent properties.

In conclusion, for an optical element, which can be either a foreground or an element of the payload optical path, the diffuse light flux density reaching the focal plane is expressed as:
\begin{equation}
FD_o (\lambda) = A_{tel} \cdot \Omega  \prod_{j=o+1}\phi_j(\lambda) \cdot \nu(\lambda) \cdot I_o(\lambda) 
\end{equation}
where $\prod_{j=o+1}\phi_j(\lambda)$ represents the product of the transmission of all the optical elements coming after the considered one, $o$, and $I_o(\lambda)$ is the optical element radiance.
In the most general case, the pixel field of view $\Omega$ is replaced by the user-defined solid angle $\omega$. For the optics box, the equation is modified as follows:
\begin{equation}
FD_{opt.box} (\lambda) = A_{tel} \cdot (\pi - \Omega) \cdot \nu(\lambda) \cdot I_{opt.box}(\lambda) 
\end{equation}
In this case, the optical transmission has been removed because there is no element between the optics box and the detector. Similarly, for the detector box:
\begin{equation}
FD_{det.box} (\lambda) = A_{tel} \cdot \pi \cdot \nu(\lambda) \cdot I_{det.box}(\lambda) 
\end{equation}

To convert the flux density to flux, two different cases need to be considered: (1) a photometer/slit-less spectrometer or (2) a spectrometer with a slit.
In the first case, the flux from the $o$-th contribution measured by each pixel is simply the integral of the flux density over all the sampled wavelengths:
\begin{equation}
F_o = \int_\lambda FD_o(\lambda) \, \textrm{d}\lambda
\end{equation}

If a slit is included in the spectrometer's optical path, the optical elements located after the slit follow the previous equation, but the elements located before the slit need to be treated in a slightly different way \citep{EChOSim, Mugnai2020, Sarkar2021}. Assuming the slit width is expressed in the number of pixels, $L$, the dispersed diffused light on the $i$-th pixel can be described with a convolution between the incoming flux $FD_o(\lambda)$ and a box car function of the size of the slit, $\Pi_{-L/2, \, L/2}(x)= \textrm{H}(x+L/2)-\textrm{H}(x-L/2)$, where $\textrm{H}(x)$ is the Heaviside step function. Then the flux from the $o$-th contribution on the $i$-th pixel is
\begin{equation}
F_{o, \, i} = \int_{\lambda} FD_o(\lambda) \Pi_{-L/2, \, L/2}(u(\lambda)-x_i) \,\textrm{d} \lambda
\end{equation}

\subsubsection{Final focal plane \label{final-focal-plane}}

\begin{figure}
\centering
\includegraphics[width=0.6\linewidth]{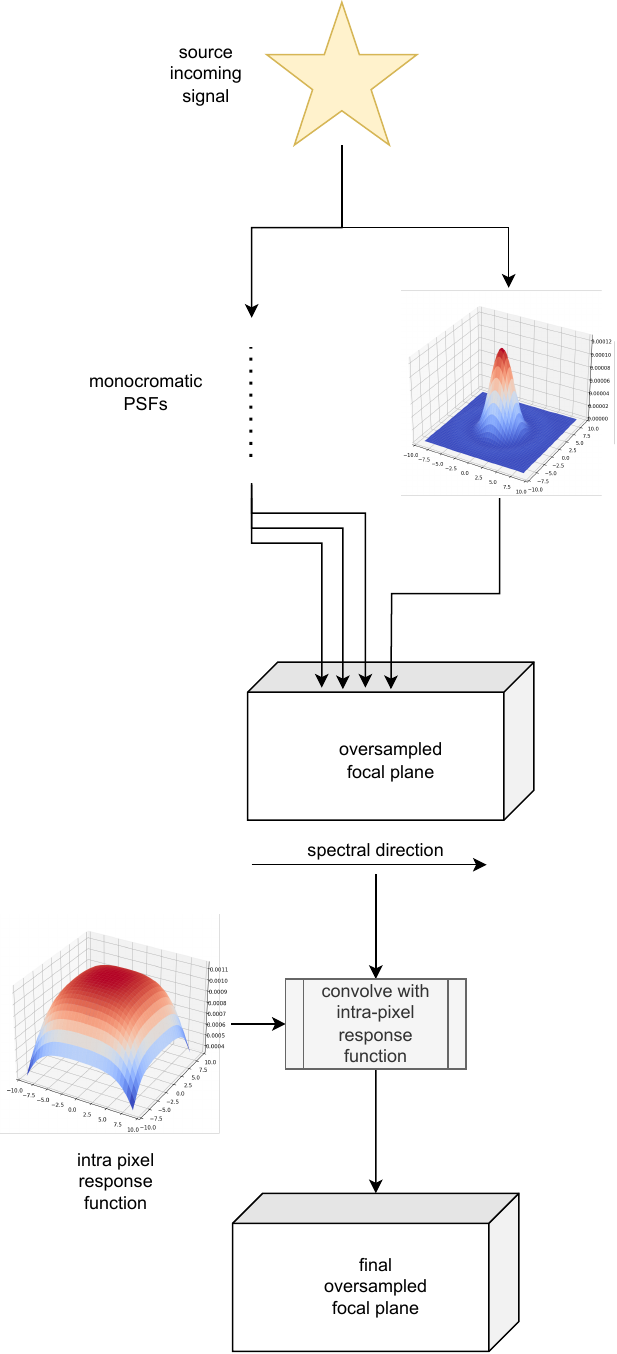}
\caption{Focal plane illumination workflow. The incoming light source is divided into monochromatic signals for each of the detector-sampled wavelengths. Each of the monochromatic signals is multiplied by the corresponding PSF. The resulting scaled PSF is added to the focal plane centered on the corresponding pixel. The resulting focal plane is then convolved with the intra-pixel response function, here represented by a default function defined as in \citet{Barron2007}.}
\label{fig:focalplanepopulation}
\end{figure}

The methodology employed by \ExoSim\ for imaging a light source on the focal plane is presented in Figure \ref{fig:focalplanepopulation}. 

As already implemented in \cite{Sarkar2021}, the illuminated focal plane is convolved with the intra-pixel response function. The result of this convolution operation is obtained for each of the $osf \times osf$ sub-pixels within a full pixel. By convolving the illuminated focal plane with the intra-pixel response function, \ExoSim\ effectively creates multiple estimates of the focal plane by shifting the original oversampled focal plane. Specifically, it generates $osf \times osf$ focal planes, each corresponding to the original focal plane shifted by $1/osf$ of a pixel. This process allows for quick computation of the jittering effect described in Sec. \ref{sub-exposure}. By default, \ExoSim\ estimates the intra-pixel response function using the prescription of \cite{Barron2007}, although this default can be replaced by the user in the configuration file. The same intra-pixel response function is applied to each pixel. Dedicated simulations have demonstrated that variations in the shape of the intra-pixel response function do not significantly affect photometric performance if the PSF is (at least) Nyquist-sampled.
After this step, the resulting oversampled focal plane comprises $osf$ focal planes, where $osf$ denotes the oversampling factor. Each focal plane represents the same image sampled as if it were shifted by $1/osf$ of the pixel size.

When considering multiple point sources, the entire process (starting from Section \ref{sec:target_source}) is repeated for each source. In such cases, the positions of the sources in the sky, along with the telescope pointing direction and the focal plane, are taken into account. These sources are then stored in what is referred to as the background focal plane. We separate these sources from the target focal plane because, in data processing, they can be considered contaminant sources. Isolating them allows us to investigate their contribution to the final signal and aids in the development of detrending algorithms or in estimating their influence. As described later in the text, these sources are subjected to the high-frequency jitter effect along with the main target.

\begin{figure}
\centering
\includegraphics[width=0.8\linewidth]{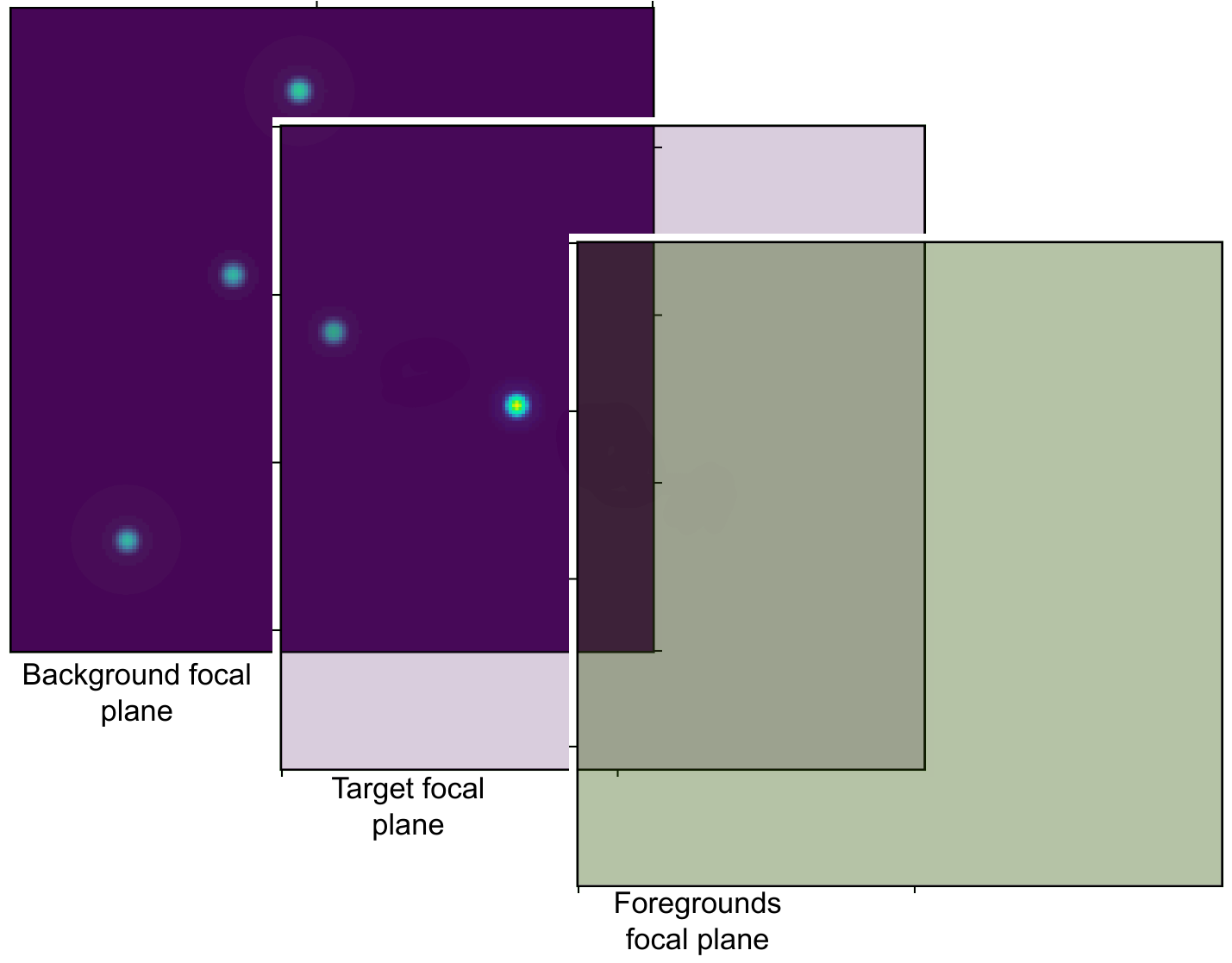}
\caption{\ExoSim\ produces three focal planes. In this example, the three focal planes are shown as three layers from left to right.  The background focal plane contains the field point sources. The target star focal plane contains only the main target to be observed with the telescope. The foreground focal plane contains the diffuse light focal plane.}
\label{fig:layers}
\end{figure}

Ultimately, \ExoSim\ generates three focal planes: the main focal plane, which contains the target source; the \textit{background focal plane}, which includes the field point sources; and the \textit{foreground focal plane}, which accommodates the diffuse light (see Fig. \ref{fig:layers}).

\ExoSim\ includes a \texttt{recipe} (\texttt{CreateFocalPlane}) that automates the focal plane creation, requiring only the input configuration file and the output file name as parameters.
The resulting focal planes are stored in an output \texttt{HDF5} file \citep{The_HDF_Group_Hierarchical_Data_Format}, along with the simulation parameters and a few key intermediate products. We stress here that the default \texttt{recipe} are intended as example procedures, and the user can change or edit them, or build their own \texttt{recipes.}   

To simulate an example of the focal plane of the Ariel mission, we utilize Point-Spread Function (PSF) simulations from \texttt{PAOS} \citep{PAOS} for the Fine Guidance Sensor (FGS) channels, namely VISPhot, FGS1, FGS2, and NIRspec. For the AIRS channels, specifically AIRS-Ch0 and AIRS-Ch1, diffraction-limited PSFs were used because i)  the telescope is designed to be diffraction-limited at 3$\mu$m and the $200 \, \mathrm{nm}$ RMS wavefront error gives diffraction-limited performance in AIRS channels and ii) this allows demonstrating the capability of \ExoSim\ to use simple Airy-shaped PSF. Also, we limit our simulation only to the Region-of-Interest (RoI) of the detector interested by the signal.

\begin{figure}	
\centering
\includegraphics[width=1\linewidth]{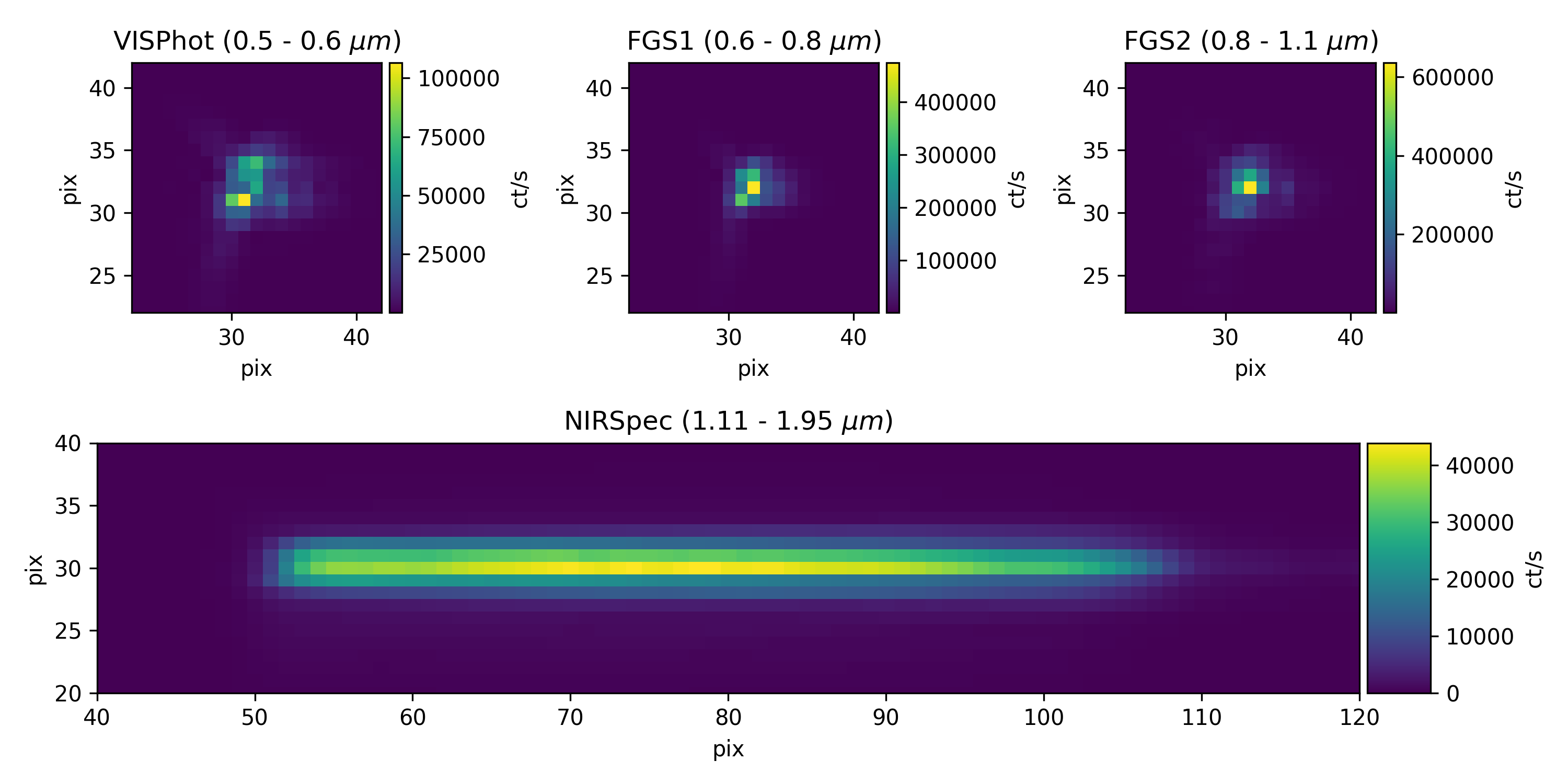}
\caption{Focal planes of the Fine Guidance Sensor (FGS) channels in the Region of Interest. The top row shows the three photometers (VISPhot, FGS1, and FGS2), while the bottom row displays the slit-less spectrometer, NIRSpec. The focal planes were generated using PAOS-simulated PSFs with a defocus of $200 \, \mathrm{nm}$ RMS. Only the signal from the star HD~209458 was included.}
\label{fig:fgs}
\end{figure}

\begin{figure*}	
\centering
\includegraphics[width=1\linewidth]{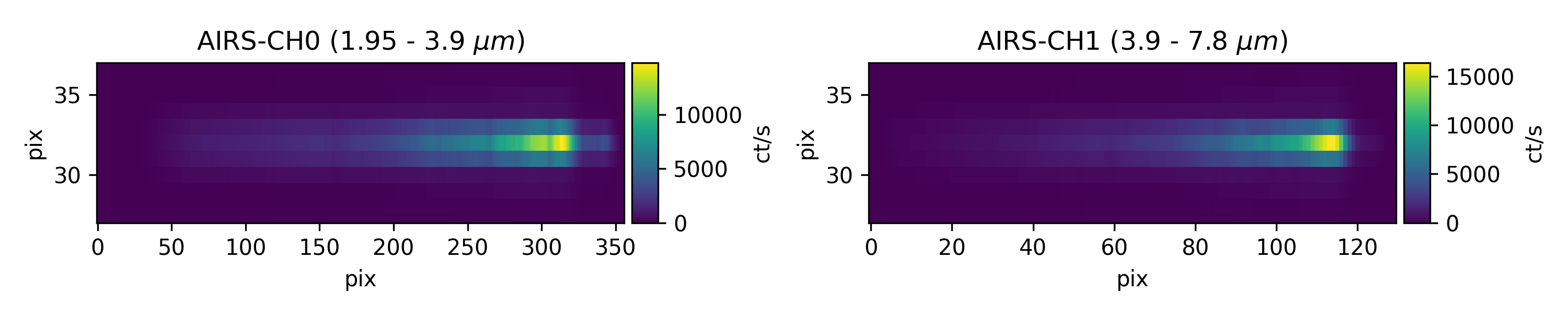}
\caption{Focal planes of the AIRS spectrometers in the Region of Interest. The left panel corresponds to Ch0, while the right panel corresponds to Ch1. The focal planes were generated using Airy PSFs and do not include contributions from diffuse light sources. Only the signal from HD 209458 was included. Note that the pixel size differs along the x and y axes.}
\label{fig:airs}
\end{figure*}

In this example, we disabled the foreground signal, which includes both the zodiacal diffuse light and the telescope self-emission, to create a controlled environment for the source signal simulated. The resulting data are presented in Figures \ref{fig:fgs} and \ref{fig:airs}.

\subsection{Sub-Exposures and Jitter}
\label{sub-exposure}

\begin{figure}	
\centering
\includegraphics[width=0.9\linewidth]{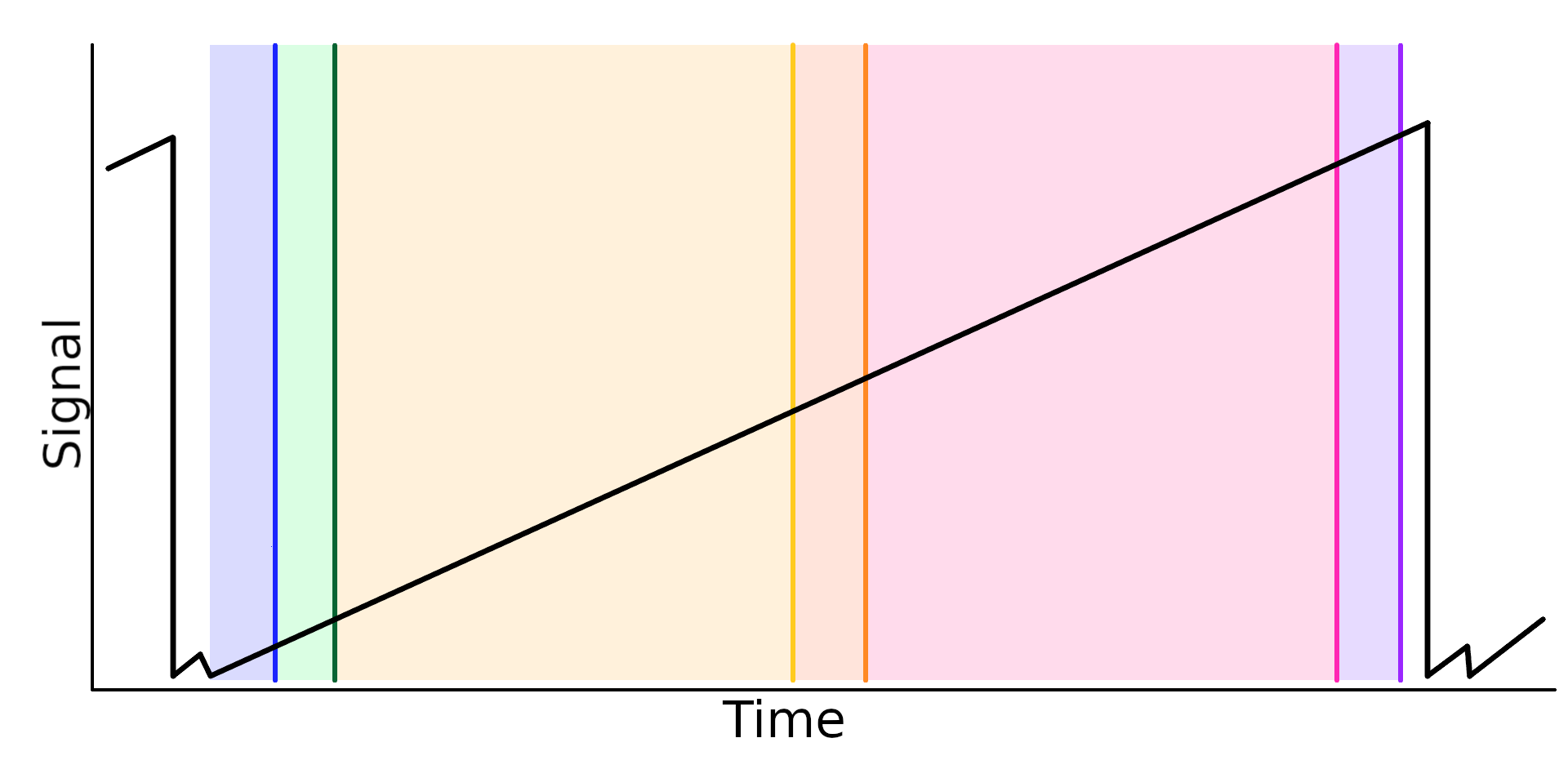}
\caption{Illustration of the detector charging ramp (black solid line) sampled by $n=3$ groups of $m=2$ NDRs (coloured lines). In this example, all NDRs between groups are discarded, storing only those specified in the readout strategy and highlighted by the solid coloured lines. Each sub-exposure (coloured regions) represents the integration of light collected starting from the end of the previous sub-exposure or the beginning of the ramp.}
\label{fig:reading-ramp}
\end{figure}

The second step of an \ExoSim\ simulation involves the charging ramp sampling. In our notation, we define ``exposure time'' as the time interval between two consecutive resets in the detector, corresponding to the integration time described in JWST papers \cite{Rauscher2007PASP}. ``Observing time'' refers to the total time spent observing the same target. Then, we define ``sub-exposure'' as the integration of the light between NDRs. Fig. \ref{fig:reading-ramp} shows the sub-exposures for a ramp read with 6 NDRs organized in 3 groups of 2 NDRs. In the figure, each sub-exposure is represented with a coloured band. The charge is integrated during each sub-exposure, contributing to each NDR, represented with solid vertical lines of different colours. Referring to the figure, the first sub-exposure (blue band) contains the light collected from the beginning of the ramp integration to the blue solid line. The second sub-exposure (green band) contains the light collected from the first NDR (blue solid line) to the green solid line (second NDR). The sub-exposures are converted into NDRs, accumulating the up-the-ramp signal in the next step of \ExoSim\ as described in Sec. \ref{NDRs}. Subsequently, the NDRs within each group are merged, simulating the on-board processing. This process assumes an instantaneous readout of the detector at each NDR. Although dedicated codes for implementing pixel sequential read-out in \ExoSim\ have been considered, for Ariel, the sequential read-out has no significant effect on the jitter representation  (see Sec. \ref{discussion:inst-readout} for details). 
This is because the jitter nuisance occurs at temporal scales that are very different from those involved in reading the detector.  
Consequently, the current version of \ExoSim\ implements an instantaneous read-out that is significantly more computationally efficient when compared to software implementations of the sequential readout scheme.

Pointing jitter occurs at a higher frequency than the detector reads: assuming that jitter is sampled at time steps $t_{hf}$, its effect needs to be included in the sub-exposures. Sub-exposures are created at the same cadence as NDRs, sampled at time steps $t_{ndr}$, which is an integer multiple of the detector reading time $t_{read}$ (the time required for the detector to read the full RoI window). As a reference, we assume $t_{lf} \sim 1 \, hr$, $t_{hf} \sim 0.01 \, s$, $t_{read} \sim 0.1 \, s$, and $t_{ndr} \sim 1 \, s$. It is important to note that $t_{ndr}$ is not fixed and depends on the reading strategy used to fit the ramp. In this example $t_{ndr} = 100 \times t_{hf}$, therefore each NDR contains 100 jitter positions. 
For another example using the same detector ($t_{read} = 0.1 \, \mathrm{s}$) and jitter timeline ($t_{hf} = 0.01 \, \mathrm{s}$), consider a Correlated Double Sampling (CDS) readout strategy. This readout mode collects 2 groups of 1 NDR each. The first NDR of the ramp is stored at $t_{ndr_0} = 0.1 \, \mathrm{s}$, and the last NDR is stored after $t_{ndr_1} = 0.9 \, \mathrm{s}$, resulting in two images. The first image is integrated for $0.1 \, \mathrm{s}$ and contains 10 jitter positions, while the second image is integrated for $1 \, \mathrm{s}$ and contains 100 jitter positions. All other NDRs that the detector would read at a cadence of $0.1 \, \mathrm{s}$ are not stored. Different possible readout strategies would allow the user to store all the NDRs, offering versatility while improving code efficiency in terms of computing time and output size.

\begin{figure}	
\centering
\includegraphics[width=1\linewidth]{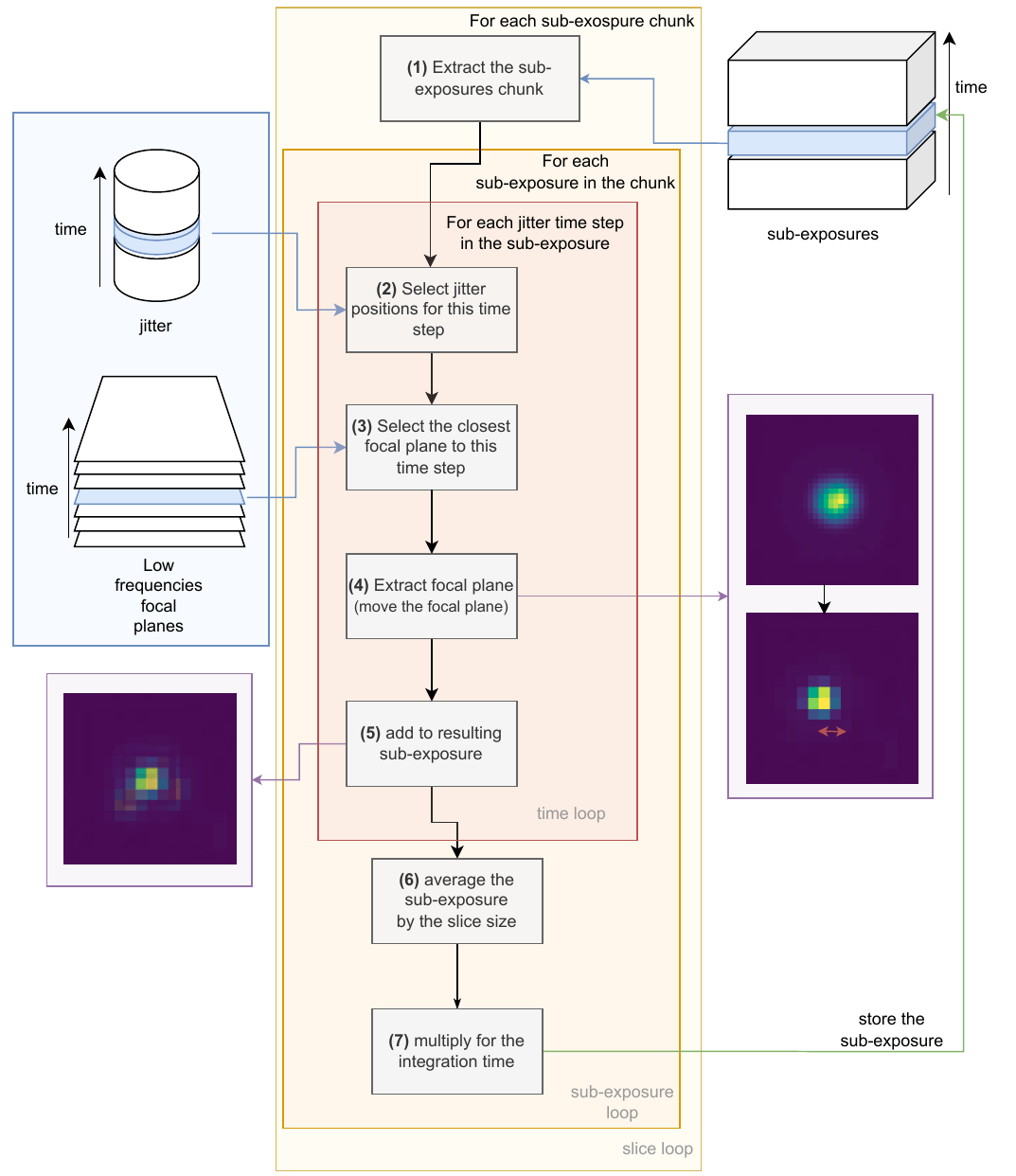}
\caption{Illustration of the chunking system in \ExoSim. First, an empty chunked output is prepared to store the sub-exposure. Then, iterations are performed over the chunks, with each iteration extracting a slice of the sub-exposure. Each sub-exposure has a set of simulation times/pointing jitter steps associated with it. For each of these time steps, the corresponding low-frequency-sampled focal plane is selected. The focal plane is shifted by the offset quantity from the oversampled focal plane. Further details about the chunking system can be found in the \ExoSim\ documentation.}
\label{fig:jittering}
\end{figure}

The nature of the simulation, which seeks to reproduce the outcome of a real observation, results in the creation of a large-volume data output. For instance, assuming a focal plane with $N \times M = 64 \times 64$ pixels, sampled with $n=3$ groups of $m=2$ NDRs as shown in Fig. \ref{fig:reading-ramp}, where each ramp is sampled every $1 \, s$ (a realistic case for bright targets), an $8 \, hr$ observation requires 28800 ramps and $n_{NDR} = 172800$ sub-exposures. Using \texttt{float64} values\footnote{We use \texttt{float64} to optimize the numerical resolution.} (64 bits = 8 bytes), the resulting data cube for the channel size is approximately $N \times M \times n_{NDR} \cdot 8 \, \text{bytes} \sim 6 \, \text{GB}$. \ExoSim\ adopts a memory-efficient strategy to deal with this demanding computation, with the introduction of the chunking system (see Fig. \ref{fig:signal_class}), which enables the creation of a large data set while optimizing memory usage (see Fig. \ref{fig:jittering}). The chunk size in megabytes can be set by the user, with the default value being $2 \, \text{MB}$. Fig. \ref{fig:jittering} reports the jitter algorithm for \ExoSim. (1) The code works with a chunk of data per time. Inside each chunk, for each sub-exposure the code (2) extracts jitter positions sampled in the sub-exposure integration time, and (3) the low-frequency sampled focal plane closest to the considered sub-exposure time; (4) for each position it takes the shifted target focal plane described in Sec. \ref{final-focal-plane} from the oversampled focal plane, and (5) adds it to the sub-exposure. Then, (6) the jittered focal planes associated with the same sub-exposure are averaged. Finally, (7) the resulting sub-exposure is multiplied by its integration times, converting from counts per second ($[ct/s]$) of the focal plane to counts ($[ct]$). Note that every pixel non-linearity effect will be applied to the data in the following step of the simulation. This process is optimized using \texttt{numba} \citep{numba} and parallel processing. The user can specify the number of parallel threads to be used in the configuration file.

Finally, \ExoSim\ supports the inclusion of astronomical effects in the simulation. Astronomical temporal effects are sampled at the same cadence of sub-exposures. The user can either define a temporal evolving source sampled at the same or higher cadence of the NDRs (see sec. \ref{sec:target_source}) or inject astronomical effects as a perturbation of a constant target source. While the first implementation may require the creation of a big data set as input for the simulation, the second strategy can suit most applications and is more computationally efficient.
To implement the latter solution, the new \ExoSim\ code includes a module for loading and applying different ``astronomical signals'' to the target source. This expands upon the ``Astroscene'' functionality of the previous \texttt{ExoSim} software \citep{Sarkar2021}, allowing users to write their own \texttt{Tasks} that describe the target astronomical events, characterized by their light-curve. The astronomical signal is then convolved with the spectral instrument line shape, to reflect the instrument response to a wavelength-dependent perturbation on the target source flux. This functionality is useful to evaluate e.g.\ the data reduction pipeline's capability to recover spectral features of a given amplitude. However, it should be noted that astronomical signals are not a core part of the framework, as \ExoSim's main aim is to reproduce complex systematic effects at the instrument level. 

The background focal plane, which contains point-like field sources, is also subject to jitter, similar to the main target focal plane. This produces a sub-exposure timeline for the background focal plane with the same shape as the target focal plane. The two sub-exposure timelines are then combined by adding the background focal plane to the target focal plane.

The foreground focal plane, which contains diffuse light, is not affected by the pointing jitter, whose impact on diffuse light is negligible. The foregrounds are therefore simply added to the sub-exposure.

\begin{figure}
\centering
\includegraphics[width=0.8\linewidth]{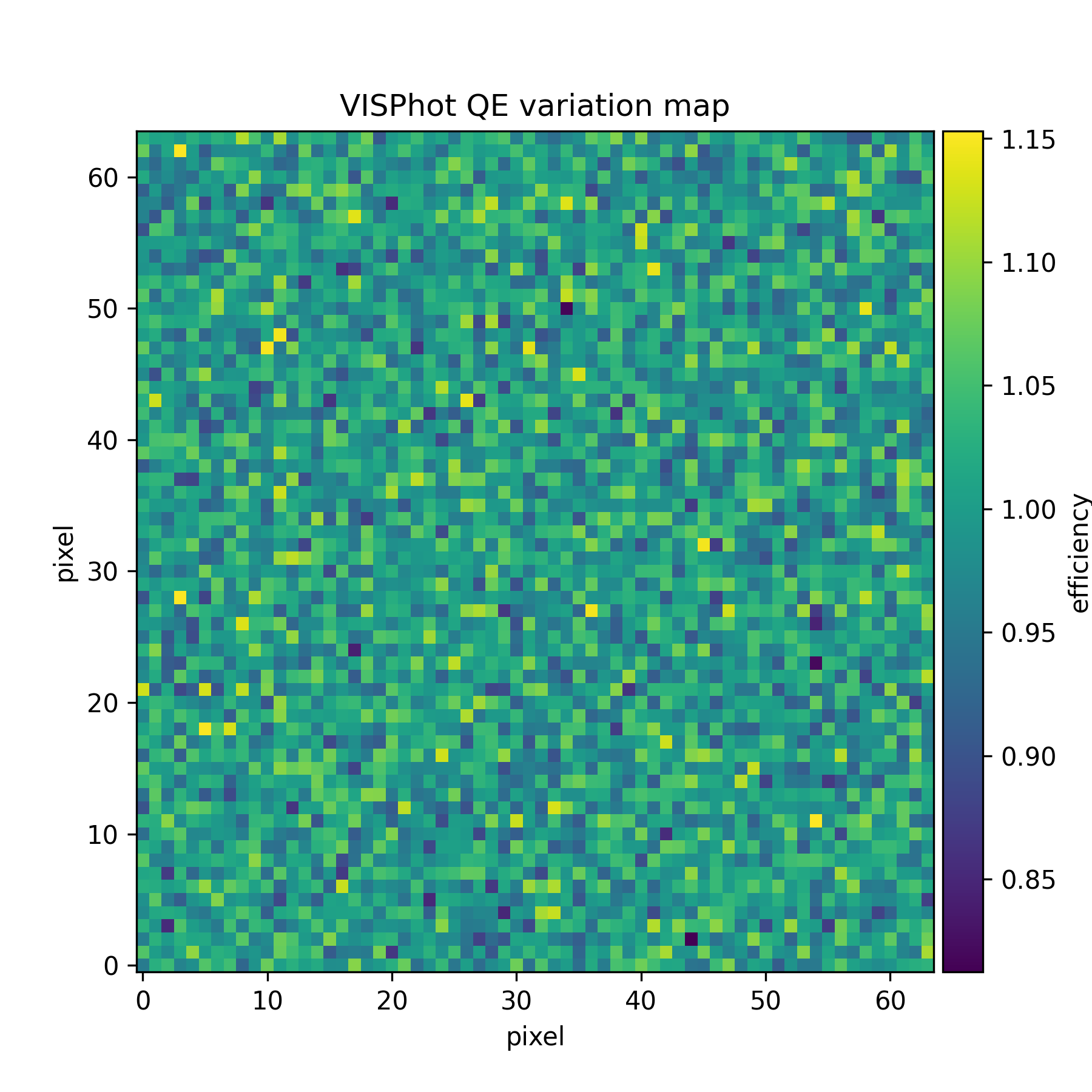}
\caption{Quantum efficiency (QE) variation map used to simulate the VISPhot channel. This map is multiplied with the focal plane after simulating the pointing jitter.}
\label{fig:qe_variation}
\end{figure}

Due to slight variations in pixel-to-pixel responsivity across the focal plane, each pixel is multiplied by a quantum efficiency variation map (see Fig. \ref{fig:qe_variation}) if provided by the user.

In a similar way to the focal planes, the creation of sub-exposures is automated through a default \texttt{recipe} (\texttt{CreateSubExposures}), which takes the configuration file and focal planes as input. The resulting sub-exposures are stored in the output \texttt{HDF5} file, along with the pointing jitter timelines in units of degrees.

\subsection{NDRs and detector noise components}
\label{NDRs}
\begin{figure}
\centering
\includegraphics[width=\linewidth]{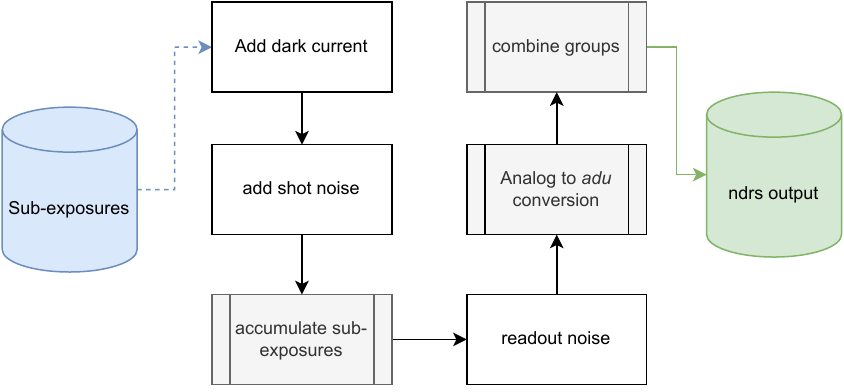}
\caption{Workflow for generating NDRs from sub-exposure. The process involves adding detector noise components to the sub-exposure and converting it into actual NDRs. For the purposes of this paper, the considered noise components are dark current, Poisson noise from photon noise, and Gaussian readout noise.}
\label{fig:ndrs}
\end{figure}

This section describes the process of generating observation NDRs in \ExoSim, along with the inclusion of detector noise components. Figure \ref{fig:ndrs} shows the schematic representation of the corresponding workflow.

Initially, \ExoSim\ incorporates the dark current into each sub-exposure, followed by the computation of shot noise. To ensure reproducibility, \ExoSim\ stores the simulation random seed and includes it in the output along with every randomly generated quantity. The user also has the option to set a specific seed value.

The sub-exposures belonging to the same ramp are accumulated to simulate up-the-ramp integration. Other detector noise components, such as kTC noise, dead and hot pixels, pixel non-linearity, persistence, cross-talk, readout noise, cosmic rays, and \textit{1/f} noise, can be included in the simulation. For the effects relevant to the Ariel space mission, a default model is already implemented in \ExoSim\ and described in the code documentation. For other effects (e.g., cross-talk and persistence), the development of a default model has been postponed, as these effects are considered negligible for the mission under study. Finally, the software is continually evolving, and additional effects, such as \textit{1/f} noise, are planned for implementation in future iterations of the simulator. Although we have developed all noise models independently, we would like to remind the reader that other dedicated software for generating noise for HxRG detectors is available, such as \citet{Rauscher2015}. However, the treatment of these effects is beyond the scope of this article.

The counts in each NDR are converted to analog-to-digital units (ADU) by casting the 64-bit floating-point numbers into n-bit integers. Since n-bit integers can only represent a maximum value, a gain conversion factor and an offset value are necessary to rescale the floating-point numbers used to represent the counts in the NDR, thereby exploiting the full dynamic range of the integer format.

To simulate different reading methods, the NDRs belonging to the same group are merged together, mimicking the MULTIACCUM reading \citep{Rauscher2007PASP, Batalha2017PASP}.

The production of NDRs is handled by the \texttt{CreateNDRs} recipe, which takes the configuration file and the pre-existing sub-exposures as inputs. The resulting NDRs are stored in a separate output \texttt{HDF5} file.

\section{Benchmark and validation}
\label{discussion} 

\subsection{Validation against ArielRad}

To validate the payload representation inside \ExoSim, we compare the photon conversion efficiency estimated by the code, which is defined as the product between the optical transmission (Fig. \ref{fig:efficiency}) and the quantum efficiency, with the one estimated by ArielRad. An automated test is included in the Ariel \ExoSim\ model that assures that the difference between the two estimates is always under $1\%$. In this study, we replicate the validation procedure presented in ArielRad \citep{Mugnai2020}, comparing the saturation time estimated for identical targets and matching payload descriptions. The results obtained with \ExoSim\ are consistent within the $5\%$ threshold defined in the ArielRad publication. Again, to ensure consistency with future code updates, an automatic test has been incorporated into the Ariel \ExoSim\ model. Similarly, automated tests have been included to automatically compare the estimated signals from the same target and implement the same payload configurations. Finally, we developed different independent data reduction pipelines specifically designed for processing \ExoSim\ data, finding differences between the \ExoSim\ estimated noise and the ArielRad noise estimation below the $1 \, \mathrm{ppm}$ level, which we consider the numerical precision of the simulation, for a $10 \, \mathrm{hr}$ observation simulation of a star representative of HD~209458, taking into account jitter noise, photon noise, and read noise.

\subsection{Instantaneous read-out validation}
\label{discussion:inst-readout}
To validate the instantaneous read-out and the jitter simulation, we compared our results to an analytical solution. We simulated a Gaussian blur on the focal plane and compared the resulting image with a Gaussian filter applied to the same initial Point Spread Function (PSF). To pass this test, any discrepancies in the amplitude and shape between the two images, if present, should be within a tolerance of $<6\%$.

To perform this validation, we generated a Gaussian PSF and simulated a Gaussian distributed pointing jitter. The obtained image was then compared to the Gaussian-filtered version of the initial PSF. Our analysis demonstrated that the differences in amplitude and shape between the two images remained below the predefined threshold of $6\%$. This validation test has been incorporated into the automatic test suite of \ExoSim to verify the consistency of successive code versions.

Alongside \ExoSim, we developed a read-out simulator, which will be presented in a separate paper. This simulator is capable of reproducing the sequential read-out of detector pixels, providing an estimate of the noise resulting from the pointing jitter when combined with different pixel read-out modes. This read-out simulator employs a flexible simulation approach, enabling instantaneous read-out, where all the pixels are read simultaneously, as is the approximation utilized by \ExoSim. We have compared the outputs of the sequential and instantaneous read-outs across various realistic jitter timelines and found that the differences are negligible compared to the photon noise anticipated for the mission. Therefore, the use of instantaneous read-out is justified and suitable for our purposes, as it aligns well with the required performance levels for the Ariel application.

\subsection{Astronomical signal validation}

To validate the astronomical signal module of \ExoSim, we simulated a primary transit observation of HD~209858~b using the default \texttt{Task} included in the package. The default \texttt{Task} in \ExoSim\ to build the transit light curve is based on the \texttt{batman-package} \cite{batman}.
This simulation focuses solely on the target star, omitting any foreground or field stars to simplify the subsequent data reduction process. The read-out mode is configured to employ a correlated double sampling method with an integration time of $3 \, s$. For the purpose of this validation, we disregard jitter effects and quantum efficiency variations between pixels, and no additional noise is introduced into the data. Because no noise effects are included, the simulated data can be assumed to represent the reduced data perfectly.

\begin{figure*}	
\centering
\includegraphics[width=.9\linewidth]{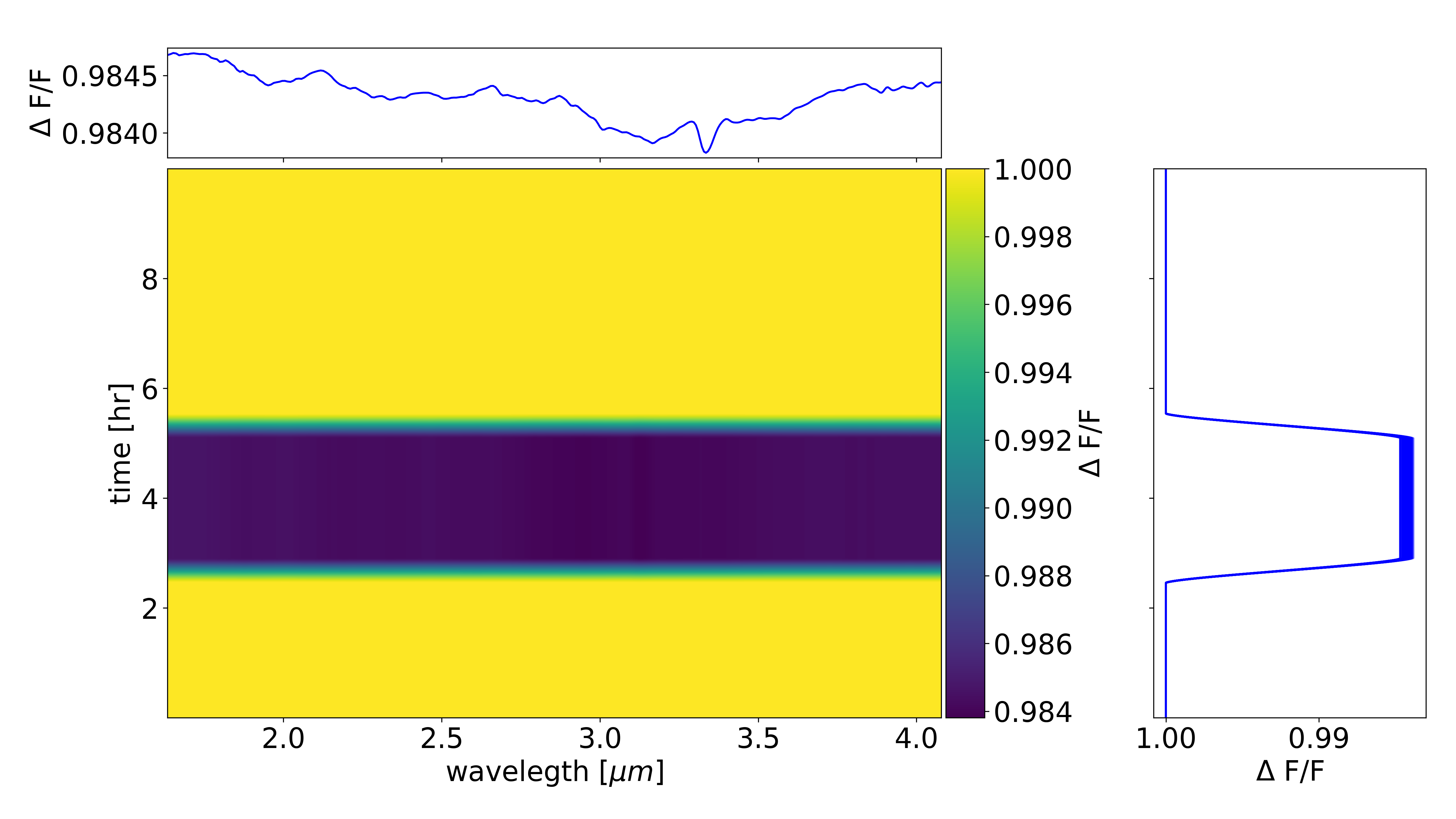}
\caption{Temporal evolution of the data modulation in AIRS-CH0 resulting from the astronomical signal simulation in \ExoSim. The images are collapsed along the spatial direction to extract the spectrum. The main panel displays the spectrum evolution over time, while the right panel presents its projection along the spectral direction and serves as a reference for the color scale of the main panel. The top panel shows the spectrum at the center of the transit.}
\label{fig:timeline_airs}
\end{figure*}

We extract the processed images from the output of \ExoSim\ for the three spectrometers and sum them along the spatial direction. Subsequently, we normalize each resulting spectrum by dividing it by the first spectrum in the time series. The resulting spectrum for AIRS-Ch0 is depicted in Figure \ref{fig:timeline_airs}. The figure shows on the central panel the timeline for each pixel column. The projection along the temporal direction is reported on the right panel, showing the transit light curves. On the top is reported the flux drop measured along the spectral direction. 

\begin{figure*}	
\centering
\includegraphics[width=.9\linewidth]{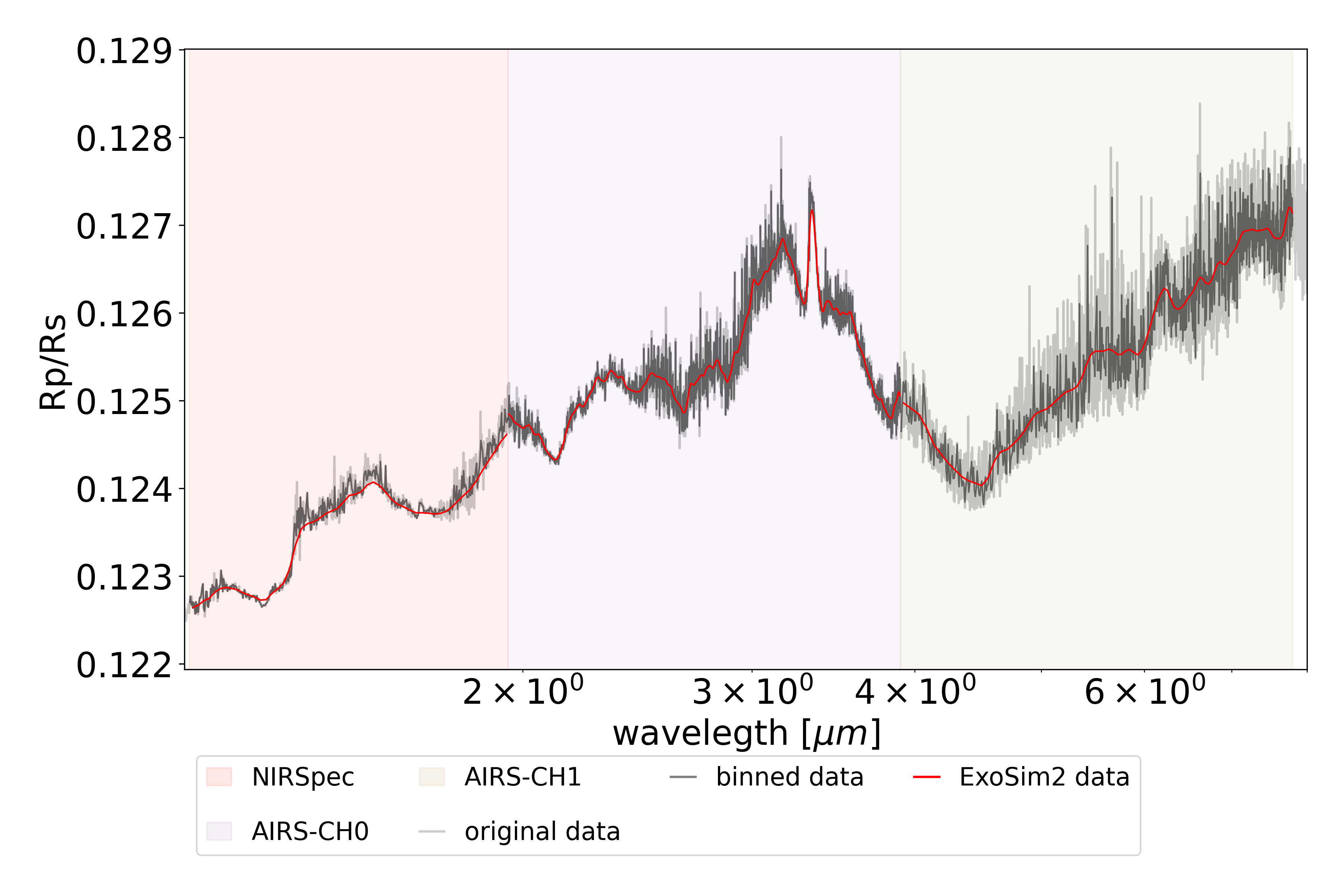}
\caption{Transmission spectrum simulated in \ExoSim. The gray curve represents the original high-resolution spectrum used as input for the astronomical signal simulation. The black curve corresponds to the same spectrum binned down to the focal plane resolution at the pixel level. The red curve represents the spectrum obtained as output from \ExoSim\ data, post-processing. The red spectrum is smoothed by the instrument line shape. The colored background indicates the spectral range of each Ariel spectrometer.}
\label{fig:spectrum}
\end{figure*}

Next, we compute the wavelength-dependent planetary radius from the flux drop, and we compare it with the input. Figure \ref{fig:spectrum} illustrates this comparison. The figure demonstrates a smoothing effect in the output spectrum when compared to the input, caused by the modulation from the instrument line shape.

% \subsection{NDRs validation}

% \subsection{other possible applications}
\subsection{Code Benchmark}

To assess the computational efficiency of the code, we conducted a series of benchmark tests. We performed 10 full simulations and computed the median time required for execution. The simulations consisted of a single source and a zodiacal foreground. We employed the Point Spread Function (PSF) computed by \texttt{PAOS} for VISPhot, FGS1, FGS2, and NIRSpec, while the Airy PSF was used for AIRS-CH0 and AIRS-CH1. The intra-pixel response function was determined using the default approach proposed by Barron et al. \cite{Barron2007}, with a diffusion length of $1.7 \, \mu \text{m}$ and no intra-pixel distance.

In the Sub-Exposure simulations, pointing jitter was sampled at a rate of $1 \, \text{kHz}$, and a correlated double sampling read-out mode was employed with an integration time of $3 \, \text{s}$. Each simulation run emulated a $10 \, \text{hr}$ observation, including a planetary transit. For NDR simulations, we considered photon noise, read noise, and dark current noise.

Although \ExoSim\ is designed to run also on standard laptops, we executed these simulations on two different server machines:
\begin{enumerate}
    \item  \texttt{Pegasus} is hosted in Cardiff University, mounting 2 \texttt{Intel(R) Core(TM) i9-10900X CPU $3.70 \, \mathrm{GHz}$ (max 4.70 $\, \mathrm{GHz}$)} with 10 cores per socket (20 threads in total)  and $256 \, \text{Gb}$ of RAM;
    \item  \texttt{Melodie} is hosted in ``Sapienza'' University of Rome, mounting 2 \texttt{Intel(R) Xeon(R) Platinum 8358 CPU 2.60 $\, \mathrm{GHz}$ (max 3.40 $\, \mathrm{GHz}$)} with 32 cores per socket (128 threads in total) and $1 \, \text{Tb}$ of RAM.
\end{enumerate} 

\begin{figure}
\centering
\includegraphics[width=1\linewidth]{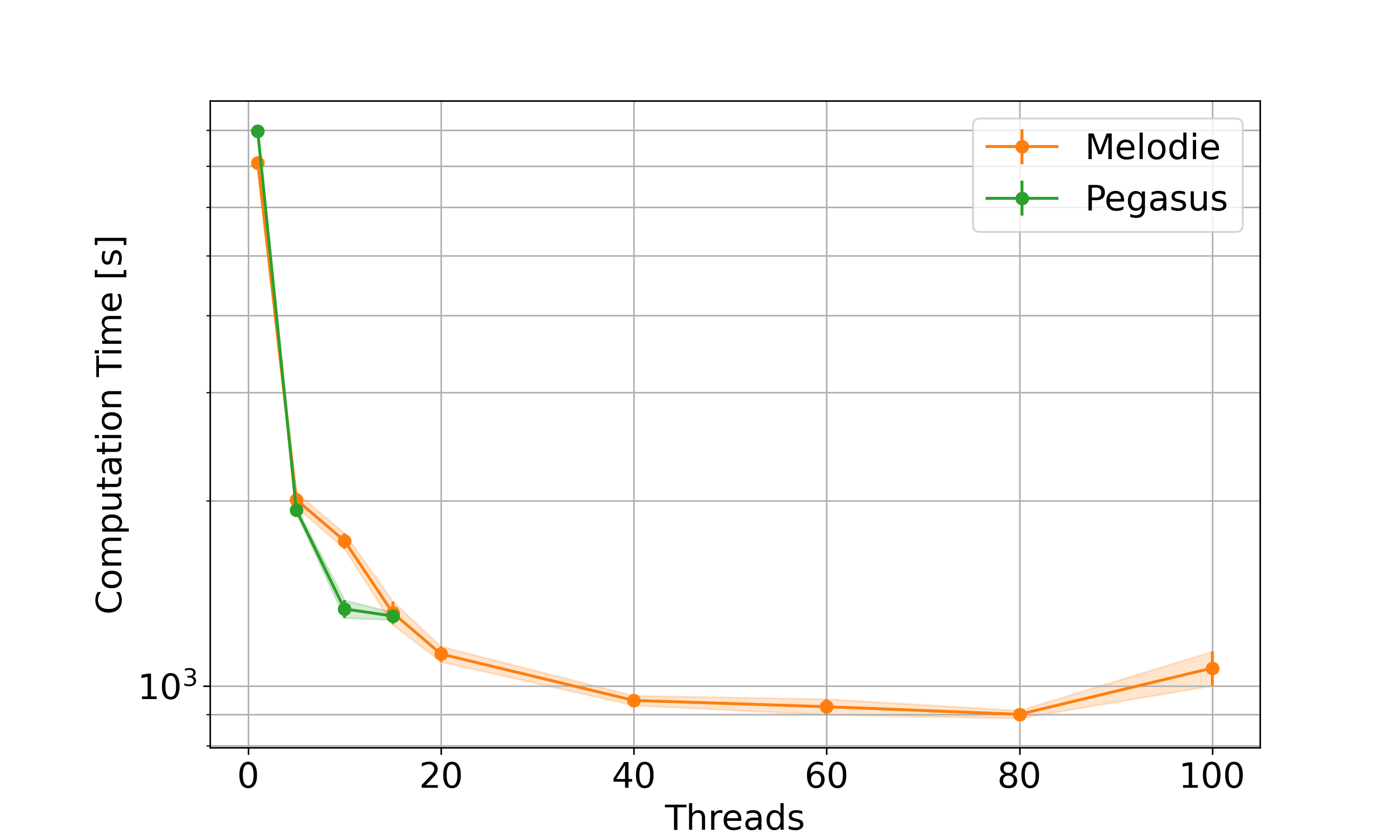}
\caption{Benchmark of \ExoSim\ performance as tested on \texttt{Pegasus} (green) and \texttt{Melodie} (orange). Data from Tab. \ref{tab:cpu_time}. Each data point is computed as the average of 10 simulations.}
\label{fig:bench}
\end{figure}

As a first test, we used 10 parallel threads (\texttt{RunConfig.n\_job = 10}), with the same chunk size (\texttt{RunConfig.chunk\_size = 5}) and random seed (\texttt{RunConfig.random\_seed = 42}).  
The median computing time was 32m:16s $\pm$ 1m:3s for \texttt{Pegasus}. \texttt{Melodie} registered a median computing time of 28m:42s $\pm$ 0m:49s.
Similarly, we tested the code running on single threads (\texttt{RunConfig.n\_job = 1}) resulting in 2h:12m:54s $\pm$ 0h:00m:30s for \texttt{Pegasus} and 1h:58m:08s $\pm$ 0h:00m:59s for \texttt{Melodie}.
Thanks to its highly parallelized algorithms, \ExoSim\ performance scales significantly with the number of threads, as shown in Tab. \ref{tab:cpu_time} and Fig. \ref{fig:bench}, while the memory used is always $< 10 \, \text{Gb}$. This data indicates that to run this simulation, the computing time limit encountered by the current version of the code is $\sim 15$ minutes. Table \ref{tab:bench_time} illustrates the relationship between computing time and three key factors: the number of stars in the field, the simulated observing time, and the detector size. All simulations were performed using 40 cores on the \texttt{Melodie} server. The HD 209458 source was duplicated to match the number of sources specified in the \texttt{.xml} file. The results indicate that including background sources increases computing time by approximately 20\%, regardless of the number of sources in the field. The data on simulated observing time show that computing time scales almost linearly with the duration of the simulation.

For varying detector sizes, simulations were conducted using only the VISPhot detector for a 10-hour observation. The results demonstrate a relatively weak dependency on the number of pixels: for instance, a detector with 64x64 pixels contains 64 times more pixels than an 8x8 detector, yet the computing time only increases by a factor of approximately 2. Similarly, a 128x128 pixel detector, which has 256 times more pixels than an 8x8 detector, results in a computing time increase of only about 6 times.

\begin{table}[]
\centering
\begin{tabular}{c|cc}
    \hline
    \# & \texttt{Pegasus} &   \texttt{Melodie} \\
    \hline
    1 &  2:12:53 (1.00) $\pm$ 0:00:30 & 1:58:08 (1.00) $\pm$ 0:00:59\\  
    5 &  0:32:12 (0.24) $\pm$ 0:0:18 & 0:33:29 (0.28) $\pm$ 0:00:55\\
    10 &  0:22:15 (0.17) $\pm$ 0:0:44 & 0:28:42 (0.24) $\pm$ 0:00:49\\
    15 &  0:21:40 (0.16) $\pm$ 0:0:18 & 0:21:56 (0.19) $\pm$ 0:00:57\\
    20 &  --                   & 0:18:48 (0.16) $\pm$ 0:00:32\\
    40 &  --                   & 0:15:48 (0.13) $\pm$ 0:00:17\\
    60 &  --                   & 0:15:26 (0.13) $\pm$ 0:00:27 \\
    40 &  --                   & 0:15:00 (0.13) $\pm$ 0:00:13\\
    100 &  --                  & 0:17:50 (0.15) $\pm$ 0:01:08 \\
    
    \hline
\end{tabular}
\caption{Relation between computing time and the number of threads. The computing time is reported in hours, minutes, and seconds. The values in parentheses indicate the computing time relative to the run using a single thread. Each estimate represents the average of 10 simulations.}

\label{tab:cpu_time}
\end{table}

\begin{table}[]
\centering
\begin{tabular}{c|c|c|c|c|c}
    \hline
    \multicolumn{2}{c|}{Number of sources} & \multicolumn{2}{c|}{Observing Time} & \multicolumn{2}{c}{Detector Size} \\
    \# & \texttt{Melodie} & [hr] & \texttt{Melodie} & Size [pix] & \texttt{Melodie} \\
    \hline
    1 & 1 (0:22:21 $\pm$ 0:00:12) & 1 &  1 (0:03:04 $\pm$ 0:00:13) & 8x8 & 1 (0:00:57 $\pm$ 0:00:01) \\
    2 & 1.198 $\pm$ 0.015 & 2 & 1.63 $\pm$ 0.16 & 16x16 & 1.05 $\pm$ 0.22 \\
    4 & 1.205 $\pm$ 0.017 & 4 & 3.01 $\pm$ 0.24 & 32x32 & 1.28 $\pm$ 0.03 \\
    8 & 1.210 $\pm$ 0.014 & 8 & 6.08 $\pm$ 0.44 & 64x64 & 2.19 $\pm$ 0.06 \\
    10 & 1.218 $\pm$ 0.014 & 10 & 7.05 $\pm$ 0.54 & 128x128 & 5.88 $\pm$ 0.25 \\
    \hline
\end{tabular}

\caption{Relationship between the number of sources, observing time, and detector size. The table shows the relative computing time for different numbers of sources (left), the simulated observing time in hours (centre), and the detector size in pixels (right). Each computing time estimate represents the average of 10 simulations, with values in parentheses indicating the actual time in hours, minutes, and seconds. All simulations were run using 40 threads.}

\label{tab:bench_time}
\end{table}

\section{Test case: the impact of jitter noise}
\label{testcase} 
Here we report an example of a use case for \ExoSim. We simulate an $8 \, hr$ observation with Ariel FGS1 of HD 209458 b, reading the detector ramp in correlated double sampling (2 NDRs per ramp) with a fixed integration time of $3 \, s$. In this simulation, we include the same configuration presented in sec. \ref{focal plane creation} and a transit light curve with a constant transit depth. We include shot noise, a constant dark current of $5 \, ct/s/pixel$, and a median read noise of $20 \, ct$. We produce two simulations, using the same noise realizations, one with jitter and one without it. 

\begin{figure*}
\centering
\includegraphics[width=1\linewidth]{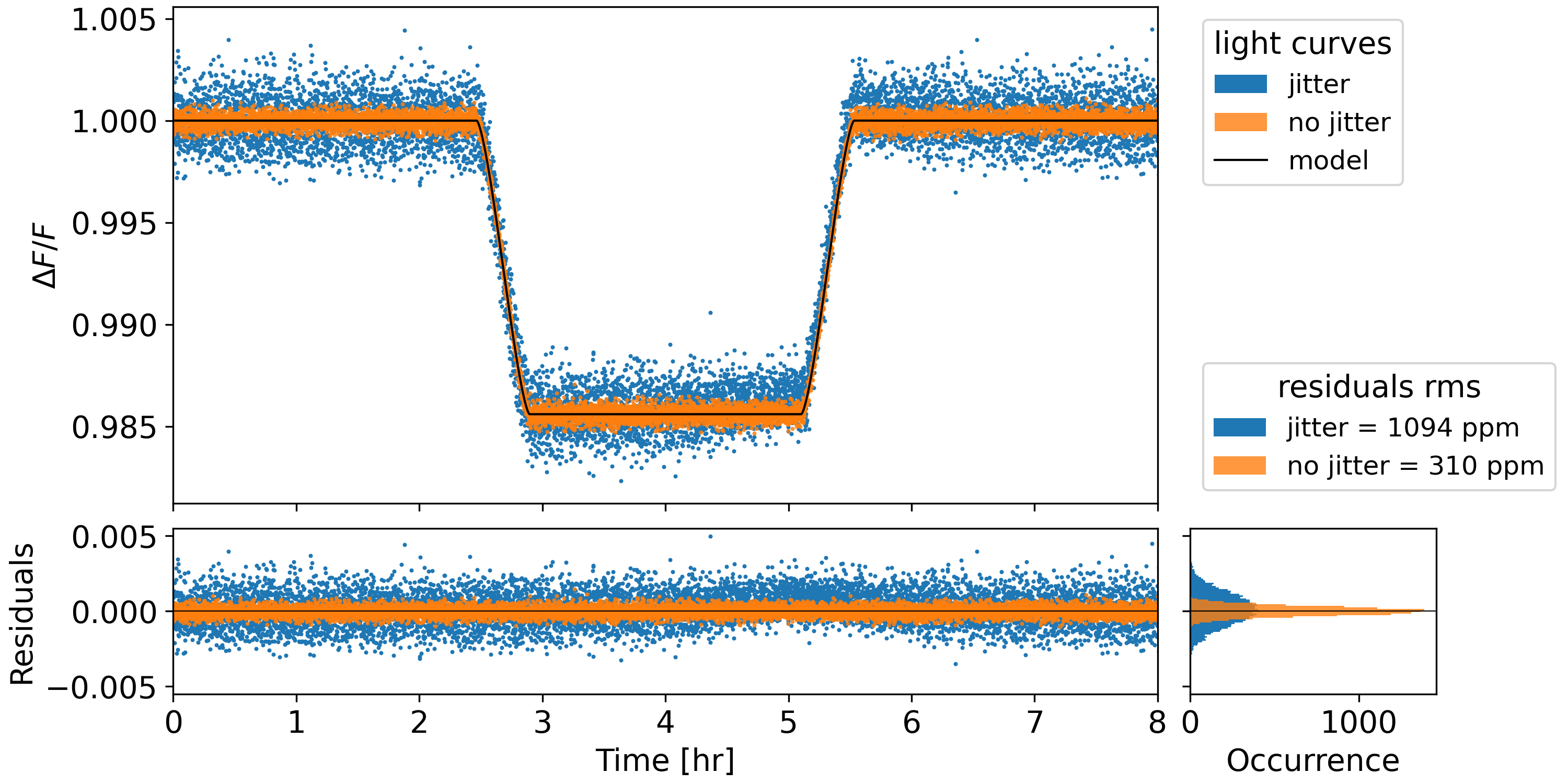}
\caption{Transit light curves (top panel) for two simulations implementing the same photon and detector noise realizations. The blue light curve includes jitter effects, while the orange curve does not. The black line is the input transit model, generated with the \texttt{batman-package}. The bottom panel shows the residuals between the light curves and the model. On the bottom right are the distributions of the residuals to highlight the differences between the jitter and the without jitter case, with the corresponding frequency histograms on the right, using the data points collapsed on the time axis. The root mean square (rms) for the residuals of each simulation is reported in the legend.}
\label{fig:use_case}
\end{figure*}

In both simulations, we perform the same data reduction. We compute the [$ct/s$] from the NDRs by subtracting the first NDR from the last one for each ramp and dividing it by the integration time. We correct the resulting images for quantum efficiency variation using a flat field correction assuming a $0.5 \%$ uncertainty in the knowledge of the quantum efficiency. We perform aperture photometry and we normalize the resulting light curve for the median of the out-of-transit data. Fig. \ref{fig:use_case} shows the resulting transit light curves. 

This test case demonstrates the impact of jitter noise on an observational dataset. Specifically, we observe a noticeable increase in the RMS of the residuals when comparing the data to the model, particularly in simulations that include jitter noise as opposed to those with only shot and detector noise, which are consistent with the expected values estimated with radiometric techniques. \ExoSim\ offers a valuable tool for developing data reduction techniques aimed at mitigating the effects of jitter noise, bringing it below the level of combined shot and detector noise.

Details about the pointing models used and the detrending technique applied to mitigate the excess 
RMS caused by jitter is described in Bocchieri et al. (2024, in preparation).

\section{Conclusion}
\label{conclusion} 

This paper presents \ExoSim, an advanced exoplanet observation simulator, designed to meet the evolving needs of the scientific community. \ExoSim\ builds upon the strengths of its predecessors \citep{EChOSim, Sarkar2021}, offering enhanced flexibility, adaptability, and a user-friendly interface. Its modular architecture, employing \texttt{Task} classes, allows for the encapsulation of simulation algorithms and functions, thereby facilitating the extensibility of the simulator to diverse instrument configurations: every piece of \ExoSim\ can be replaced by a user-defined function. \ExoSim\ comes with a comprehensive step-by-step how-to-use guide, working examples and API documentation to guide and train the user as well as the developer who wishes to contribute to the code. The software also includes a set of tools to produce examples of calibration data and estimate quantities needed for the simulation.  

\ExoSim's three-step workflow for code execution, involving the creation of focal planes, production of Sub-Exposure blocks, and generation of non-destructive reads (NDRs), optimizes the time and computational resources. This allows users to explore different configurations without the need to re-run the entire simulation.

The extensive validation of \ExoSim\ against other tools, such as ArielRad, underscores its reliability and accuracy. The simulator has demonstrated consistency in estimating photon conversion efficiency, saturation time, and signal generation. Furthermore, its validation for instantaneous read-out and jitter simulation, as well as for astronomical signal representation, attests to its robustness and versatility.

Originally designed for supporting the preparation of Ariel, \ExoSim\ has undergone rigorous testing against known systematics for the space mission, enabling predictions of their impact on the final measurements and facilitating the prototyping of data reduction pipelines. \ExoSim's versatility and user-friendliness help it transcend its initial use case, and currently, it is being applied to the EXCITE stratospheric balloon-borne mission. The ability of \ExoSim\ to simulate the physics encapsulated in user-defined models opens avenues for other observatories, such as the JWST, to be modeled within its framework. The ensuing simulated data, when combined with actual measurements, serves to assess the ability of data reduction pipelines to correct known systematics, and evaluate the presence of bias in the estimators.
Current limitations of the \ExoSim\ code stem from the need to implement new noise or detector models to accommodate effects that were not initially considered relevant for Ariel, such as detector persistence, or were postponed due to time constraints, such as 1/f noise. Other known limitations include the high-frequency module's inability to simulate slow pointing drift, as it was originally developed for fast random jitter. This limitation can be addressed by developing a dedicated drift model or enhancing the jitter model with an intermediate step interpolator. Such enhancements are planned and will be included in the next code releases. Additionally, users are encouraged to implement custom models following the comprehensive guide in the documentation, which also covers how to customise the \ExoSim\ code.

In conclusion, \ExoSim\ represents a significant advancement in the field of exoplanet observation simulation. Its design principles and validation results underscore its potential as a valuable resource for researchers. As we continue to explore the vast expanse of the cosmos to study exoplanets, tools like \ExoSim\ will play a pivotal role in shaping our understanding of these distant worlds. Ultimately, such tools are irreplaceable in understanding the biases of estimators and accurately interpret the data we collect.

Future work will focus on further enhancing the capabilities of \ExoSim, ensuring it remains at the forefront of exoplanet observation simulation. As we strive to refine our understanding of the true nature of exoplanets, \ExoSim\ will continue to serve as an indispensable asset in our scientific arsenal.

\backmatter

% \bmhead{Supplementary information}

% If your article has accompanying supplementary file/s please state so here. 

% Authors reporting data from electrophoretic gels and blots should supply the full unprocessed scans for key as part of their Supplementary information. This may be requested by the editorial team/s if it is missing.

% Please refer to Journal-level guidance for any specific requirements.

\bmhead{Acknowledgments}

This work has been supported by ASI grant n. 2021.5.HH.O. 
This work was carried out in the context of Ariel's \textit{Simulators Software, Management, and Documentation (S2MD)} Working Group. The authors thank all participants for their constant comments and suggestions.

\section*{Declarations}

\bmhead{Funding Declaration}
The authors acknowledge that this work has been supported by the ASI grant n. 2021.5.HH.0.

\bmhead{Competing Interest}
The authors declare they have no conflict of interest.

\bmhead{Author Contribution}
L.V.M. wrote the main manuscript text, prepared the figures, and developed the code for the simulator. E.P. provided guidance and oversight during the code development process. A.B. assisted in both the development and testing phases of the code. A.L. and A.P. contributed to the testing and validation of the simulator. All authors reviewed and approved the final manuscript for submission.

\bmhead{Data Availability}

The data underlying the research results in this article primarily stem from mission parameters adopted from the B2-phase of the Ariel mission. These parameters were crucial in shaping the payload configurations used in our study. To ensure reproducibility while maintaining the confidentiality of sensitive information, these payload configurations are stored in a dedicated GitHub repository. Access to this repository is restricted to individuals with key responsibilities within the Ariel mission team.

For further information on data access, interested parties may contact the corresponding author or the Ariel mission's data management team.

\bibliography{refs}% common bib file

\end{document}